\input ./style/arxiv-general.cfg
\documentclass[aoas,MSNbibl,nameyear,dvips]{arximspdf}
\makeatletter
   \@ifpackageloaded{graphicx}{}{\usepackage{graphicx}}
\makeatother
\usepackage{algpseudocode,algorithm}

\doi{10.1214/15-AOAS857}
\volume{9}
\issue{3}
\pubyear{2015}
\firstpage{1671}
\lastpage{1705}
\docsubty{FLA}

\makeatletter
\def\overset{\stackrel}
\newcommand{\eqref}[1]{(\ref{#1})}
\newcommand{\E}{\operatorname{E}}
\newcommand{\argmax}{\mathop{\operatorname{arg max} }}
\algnewcommand\algorithmicto{\textbf{to}}
\renewcommand{\algorithmicrequire}{\textbf{Input:}}
\renewcommand{\algorithmicensure}{\textbf{Output:}}
\makeatother

\begin{document}
\begin{frontmatter}

\title{Statistical unfolding of elementary particle spectra: Empirical
Bayes estimation and bias-corrected uncertainty quantification}
\runtitle{Statistical unfolding of elementary particle spectra}

\begin{aug}
\author[A]{\fnms{Mikael}~\snm{Kuusela}\thanksref{T1,TL}\ead[label=e1]{mikael.kuusela@epfl.ch}}
\and
\author[A]{\fnms{Victor M.} \snm{Panaretos}\corref{}\thanksref{T1}\ead[label=e2]{victor.panaretos@epfl.ch}}
\runauthor{M. Kuusela and V.~M. Panaretos}
\address[A]{Section de Math\'{e}matiques\\
\'{E}cole Polytechnique F\'{e}d\'{e}rale de Lausanne\\
EPFL Station 8, 1015 Lausanne \\
Switzerland \\
\printead{e1}\\
\phantom{E-mail:\ }\printead*{e2}}
\affiliation{\'{E}cole Polytechnique F\'{e}d\'{e}rale de Lausanne}

\thankstext{T1}{Supported in part by a Swiss National Science
Foundation grant.}
\thankstext{TL}{Supported in part by a grant from the Helsinki
Institute of Physics.}
\end{aug}

%
\received{\smonth{1} \syear{2014}}
%
\revised{\smonth{7} \syear{2015}}

%
\begin{abstract}
We consider the high energy physics unfolding problem where the goal is
to estimate the spectrum of elementary particles given observations
distorted by the limited resolution of a particle detector. This
important statistical inverse problem arising in data analysis at the
Large Hadron Collider at CERN consists in estimating the intensity
function of an indirectly observed Poisson point process. Unfolding
typically proceeds in two steps: one first produces a regularized point
estimate of the unknown intensity and then uses the variability of this
estimator to form frequentist confidence intervals that quantify the
uncertainty of the solution. In this paper, we propose forming the
point estimate using empirical Bayes estimation which enables a
data-driven choice of the regularization strength through marginal
maximum likelihood estimation. Observing that neither Bayesian credible
intervals nor standard bootstrap confidence intervals succeed in
achieving good frequentist coverage in this problem due to the inherent
bias of the regularized point estimate, we introduce an iteratively
bias-corrected bootstrap technique for constructing improved confidence
intervals. We show using simulations that this enables us to achieve
nearly nominal frequentist coverage with only a modest increase in
interval length. The proposed methodology is applied to unfolding the
$Z$~boson invariant mass spectrum as measured in the CMS experiment at
the Large Hadron Collider.
\end{abstract}

%
\end{frontmatter}

\section{Introduction} \label{sec:intro}

This paper studies a generalized linear inverse problem [\citet
{Bochkina2013}], called the \emph{unfolding problem} [\citet
{Prosper2011,Cowan1998,Blobel2013}], arising in data analysis at the
Large Hadron
Collider (LHC) at CERN, the European Organization for Nuclear Research.
The LHC is the world's largest and most powerful particle accelerator.
It collides two beams of protons in order to study the properties and
interactions of elementary particles produced in such collisions. The
trajectories and energies of these particles are recorded using
gigantic underground particle detectors and the vast amounts of data
produced by these experiments are analyzed in order to draw conclusions
about fundamental laws of physics. Due to their complex structure and
huge quantity, the analysis of these data poses significant statistical
and computational challenges.

Experimental high energy physicists use the term ``unfolding'' to refer
to correcting the distributions measured at the LHC for the limited
resolution of the particle detectors. Let $X$ be some physical quantity
of interest measured in the detector. This could, for example, be the
energy, mass or production angle of a particle. Due to the noise
induced by the detector, we are only able to observe a stochastically
\emph{smeared} or \emph{folded} version $Y$ of this quantity. As a
result, the observed distribution of $Y$ is a ``blurred'' version of
the true, physical distribution of $X$ and the task is to use the
observed values of $Y$ to estimate the distribution of~$X$. Each year,
the experimental collaborations working with LHC data produce dozens of
physics results that make use of unfolding. Recent examples include
studies of the characteristics of jets [\citet{CMS2012Jets}], the
transverse momentum distribution of $W$ bosons [\citet{ATLAS2012W}] and
charge asymmetry in top-quark pair production [\citet{CMS2012Top}], to
name a few.

The main challenge in unfolding is the ill-posedness of the problem in
the sense that a simple inversion of the forward mapping from the true
space into the smeared space is unstable with respect to small
perturbations of the data [\citet{Engl2000,Kaipio2005,Panaretos2011}].
As such, the trivial maximum likelihood solution of the problem often
exhibits spurious high-frequency oscillations. These oscillations can
be tamed by regularizing the problem, which is done by taking advantage
of additional a priori knowledge about plausible solutions.

An additional complication is the non-Gaussianity of the data which
follows from the fact that both the true and the smeared observations
are realizations of two interrelated Poisson point processes, which we
denote by $M$ and $N$, respectively. As such, unfolding is an example
of a \emph{Poisson inverse problem} [\citet{Antoniadis2006,Reiss1993}],
where the intensity function $f$ of the true process $M$ is related to
the intensity function $g$ of the smeared process $N$ via a Fredholm
integral operator $K$, that is, $g = Kf$, where $K$ represent the
response of the detector. The task at hand is then to estimate and make
inferences about the true intensity $f$ given a single observation of
the smeared process $N$. Due to the Poisson nature of the data, many
standard techniques based on a Gaussian likelihood function, such as
Tikhonov regularization, are only approximately valid for unfolding.
Furthermore, estimators properly taking into account the Poisson
distribution of the observations are rarely available in a closed form,
making the problem computationally challenging.

At present, the unfolding methodology used in LHC data analysis is not
well established [\citet{Lyons2011}]. The two main approaches are the
expectation--maximization (EM) algorithm with an early stopping [\citet
{DAgostini1995,Vardi1985,Lucy1974,Richardson1972}] and a certain
variant of Tikhonov regularization [\citet{Hoecker1996}]. In high energy
physics (HEP) terminology, the former is called the \emph{D'Agostini
iteration} and the latter, somewhat misleadingly, \emph{SVD unfolding}
(with SVD referring to the singular value decomposition). In addition,
a HEP-specific heuristic, called \emph{bin-by-bin unfolding}, which
provably accounts for smearing effects incorrectly through a
multiplicative efficiency correction, has been widely used. Recently,
\citet{Choudalakis2012} proposed a Bayesian solution to the problem,
but this seems to have seldom been used in practice thus far.

The main problem with the D'Agostini iteration is that it is difficult
to give a physical interpretation to the regularization imposed by
early stopping of the iteration. SVD unfolding, on the other hand,
ignores the Poisson nature of the observations and does not enforce the
positivity of the solution. Furthermore, both of these methods suffer
from not dealing with two significant issues satisfactorily: (1) the
choice of the regularization strength and (2) quantification of the
uncertainty of the solution. The delicate problem of choosing the
regularization strength is handled in most LHC analyses using
nonstandard heuristics or, in the worst-case scenario, by simply fixing
some value ``by hand.'' When quantifying the uncertainty of the
unfolded spectrum, the analysts form approximate frequentist confidence
intervals using simple error propagation, but little is known about the
coverage properties of these intervals.

In this paper, we propose new statistical methodology aimed at
addressing the two above-mentioned issues in a more satisfactory
manner. Our main methodological contributions are as follows:
\begin{longlist}[1.]
\item[1.] Empirical Bayes selection of the regularization parameter using a
Monte Carlo expectation--maximization algorithm [\citet{Casella2001},
Geman and McClure (\citeyear{Geman1985,Geman1987}), \citet{Saquib1998}];
\item[2.] Frequentist uncertainty quantification using a combination of an
iterative bias-correction procedure [\citet{Kuk1995,Goldstein1996}] and
bootstrap percentile intervals [\citet{Efron1993,Davison1997}].
\end{longlist}
To the best of our knowledge, neither of these techniques has been
previously used to solve the HEP unfolding problem. Our framework also
properly takes into account the Poisson distribution of the
observations, enforces the positivity constraint of the unfolded
spectrum and imposes a curvature penalty on the solution with a clear
physical interpretation.

It is helpful to think of the unfolding problem as consisting of two
separate, but related, subproblems: one of point estimation and the
other of uncertainty quantification. We follow the standard practice of
first constructing a point estimate of the unknown intensity and then
using the variability of this point estimate to form frequentist
confidence intervals. For the point estimation part, the main challenge
is to decide how to regularize the ill-posed problem and, in
particular, how to choose the regularization strength. Classical,
well-known techniques for making this choice include the Morozov
discrepancy principle [\citet{Morozov1966}] and cross-validation
[\citet
{Stone1974}]. \citet{Bradsley2009} study these techniques in the
context of Poisson inverse problems, while \citet{Veklerov1987} provide
an alternative approach based on statistical hypothesis testing. From a
Bayesian perspective, the problem can be addressed using a Bayesian
hierarchical model [\citet{Kaipio2005}] which necessitates the choice of
a hyperprior for the regularization parameter. On the other hand,
empirical Bayes selection of the regularization parameter using the
marginal likelihood, which has the key advantage of not requiring the
specification of a hyperprior, has received relatively less attention
in the inverse problems literature. However, in many other fields of
statistical inference, such as semiparametric regression [\citet
{Wood2011,Ruppert2003}, Section~5.2], neural networks
[\citet{Bishop2006}, Sections~3.5 and 5.7.2] and
Gaussian processes [\citet{Rasmussen2006}, Chapter~5], the use of
empirical Bayes techniques has become
part of standard practice. The approach we follow bears similarities to
that of \citet{Saquib1998}, where the marginal maximum likelihood
estimator is used to select the regularization parameter in tomographic
image reconstruction with Poisson data.

Once we have formed an empirical Bayes point estimate of the unknown
intensity function, we would like to use the variability of this point
estimator to quantify the uncertainty of the solution. In high energy
physics, frequentist confidence statements are generally preferred over
Bayesian alternatives [\citet{Lyons2013}] and we would hence like to
obtain confidence intervals with good frequentist coverage properties.
To achieve this, the main challenge comes from the bias that is present
in the point estimate in order to regularize the otherwise ill-posed
problem. We show using simulations that this bias can seriously degrade
the coverage probability of both Bayesian credible intervals and
standard bootstrap confidence intervals. We propose to remedy this
problem by employing an iterative bias-correction technique [\citet
{Kuk1995,Goldstein1996}] and then using the variability of the
bias-corrected point estimate to form the confidence intervals. Quite
remarkably, our simulation results indicate that such a technique
achieves close-to-nominal coverage with only a modest increase in the
length of the intervals.

The paper is structured as follows. Section~\ref{sec:physics} gives a
brief overview of the LHC detectors and explains the role that
unfolding plays in these experiments. We then formulate in Section~\ref
{sec:formulation} a forward model for the unfolding problem using
Poisson point processes. The proposed statistical methodology is
explained in detail in Section~\ref{sec:EBU} which forms the backbone
of this paper. This is followed by simulation studies in Section~\ref
{sec:simulations} and a real-world data analysis scenario in
Section~\ref{sec:Zboson} which consists of unfolding of the $Z$ boson invariant
mass spectrum measured in the CMS experiment at the LHC. We close the
paper with some concluding remarks in Section~\ref{sec:discConc}. We
also invite the reader to consult the online supplement [\citet
{Kuusela2015}] which provides further simulation results and some
technical details.

\section{LHC experiments and unfolding}

\label{sec:physics}

\subsection{Overview of LHC experiments} \label{sec:physicsDetectors}

The Large Hadron Collider is a\break 27~km long circular
proton--proton collider located in an underground tunnel at CERN in
Geneva, Switzerland.
With proton--proton collisions of up to\break 13~TeV\setcounter{footnote}{2}\footnote{The electron
volt, $\mathrm{eV}$, is the customary unit of energy used in particle
physics, $1\ \mathrm{eV} \approx1.6 \cdot10^{-19}\ \mathrm{J}$.}
center-of-mass energy, the LHC is the world's most powerful particle
accelerator. The protons are accelerated in bunches of billions of
particles and bunches moving in opposite directions are led to collide
at the center of four gigantic particle detectors called ALICE, ATLAS,
CMS and LHCb. In the LHC Run 1 configuration, these bunches collided
every 50 ns at the heart of the detectors, resulting in some 20 million
collisions per second in each detector, out of which the few hundred
most interesting ones were stored for further analysis. In LHC Run 2,
which started in June 2015, the collision rate is likely to be even higher.

Out of the four detectors, ATLAS and CMS are multipurpose experiments
capable of performing a large variety of physics analyses ranging from
the discovery of the Higgs boson to precision studies of quantum
chromodynamics. The other two detectors, ALICE and LHCb, specialize in
studies of lead-ion collisions and $b$-hadrons, respectively. In what
follows, we focus on describing the data collection and analysis in the
CMS experiment, which is also the source of the data of our unfolding
demonstration in Section~\ref{sec:Zboson}, but similar principles also
apply to ATLAS and, to some extent, to other high energy physics experiments.

The CMS experiment [\citet{CMS2008}], an acronym for Compact Muon
Solenoid, is situated in an underground cavern along the LHC ring near
the village of Cessy, France. The detector, weighing a total of 12,500~tons, has a cylindrical shape with a diameter of 14.6~m and a
length of 21.6~m. The construction, operation and data analysis of the
experiment is conducted by an international collaboration of over 4000
scientists, engineers and technicians. When two protons collide at the
center of CMS, their energy is transformed into matter in the form of
new particles. A small fraction of these particles are exotic,
short-lived particles, such as the Higgs boson or the top quark, which
are at the center of the scientific interest of the high energy physics
community. Such particles decay almost instantly into more familiar,
stable particles, such as electrons, muons or photons. Using various
subdetectors, the energies and trajectories of these particles are
recorded in order to study the properties and interactions of the
exotic particles created in the collision.

%
\begin{figure}[b]

\includegraphics{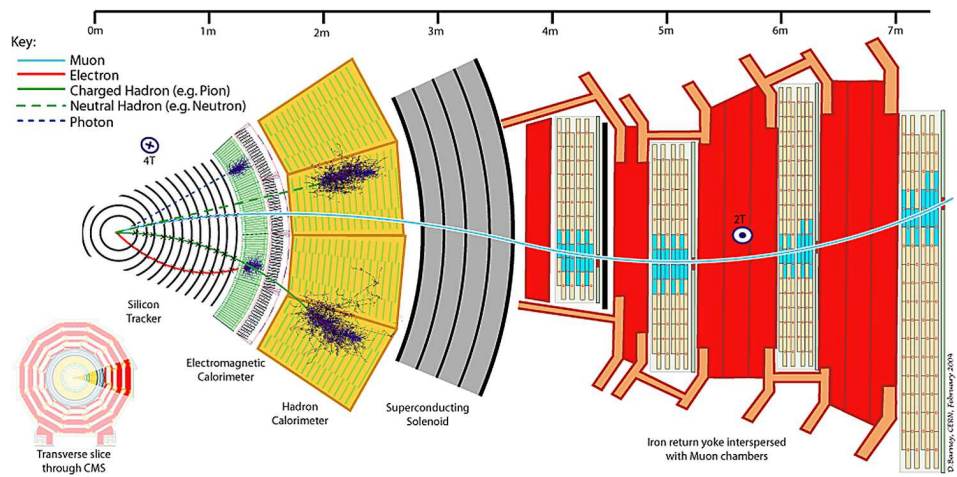}

\caption{Illustration of the detection of particles at the CMS
experiment [\citet{Barney2004}]. Each type of particle leaves its
characteristic trace in the various subdetectors of the experiment.
This enables identification of different particles as well as the
measurement of their energies and trajectories. Copyright: CERN, for
the benefit of the CMS Collaboration.}
\label{fig:CMS_slice}
\end{figure}

The layout of the CMS detector is illustrated in Figure~\ref
{fig:CMS_slice}. The detector is immersed in a 3.8~T magnetic field
created using a superconducting solenoid magnet. This magnetic field
bends the trajectory of any charged particle traversing the detector.
This enables the measurement of the particle's momentum, since the
higher the momentum, the less the particle's trajectory is bent.

CMS consists of three layers of subdetectors: the tracker, the
calorimeters and the muon detectors. The innermost detector is the
silicon tracker, which consists of an inner layer of pixel detectors
and an outer layer of microstrip detectors. When a charged particle
passes through these semiconducting detectors, it leaves behind
electron--hole pairs, and hence creates an electric signal. These
signals are combined into a particle track using a Kalman filter in
order to reconstruct the trajectory of the particle.

The next layer of detectors are the calorimeters, which are devices for
measuring the energies of particles. The CMS calorimeter system is
divided into an electromagnetic calorimeter (ECAL) and a hadron
calorimeter (HCAL). Both of these devices are based on the same general
principle: they are made of extremely dense materials with the aim of
stopping the particles passing through. In the process, a portion of
the energy of these particles is converted into light in a
scintillating material and the amount of light, which depends on the
energy of the incoming particle, is measured using photodetectors
inside the calorimeters. The ECAL measures the energy of particles that
interact mostly via the electromagnetic interaction, in other words,
electrons, positrons and photons. The HCAL, on the other hand, measures
the energies of hadrons, that is, particles composed of quarks. These
include, for example, protons, neutrons and pions. The HCAL is also
instrumental in measuring the energies of jets, that is, collimated
streams of hadrons produced by quarks and gluons, and in detecting the
so-called missing transverse energy, an energy imbalance caused by
noninteracting particles, such as neutrinos, escaping the detector.

The outermost layer of CMS consists of muon detectors, whose task is to
identify muons and measure their momenta. Accurate detection of muons
was of central importance in the design of CMS since muons provide a
clean signature for many exciting physics processes. This is because
there is a very low probability for other particles, with the exception
of noninteracting neutrinos, to penetrate through the CMS calorimeter
system. For example, the four-muon decay channel played an important
role in the discovery of the Higgs boson at CMS [\citet{CMS2012Higgs}].

The information of all CMS subdetectors is combined [\citet{CMS2009}] to
identify the stable particles, that is, muons, electrons, positrons,
photons and various types of hadrons, produced in each collision event;
see Figure~\ref{fig:CMS_slice}. For example, a muon will leave a track
in both the silicon tracker and the muon chamber, while a photon
produces a signal in the ECAL without an associated track in the
tracker. The information of these individual particles is then used to
reconstruct higher-level physics objects, such as jets or missing
transverse energy.

\subsection{Unfolding in LHC data analysis} \label{sec:physicsUnfolding}

The need for unfolding arises because any quantity measured at the LHC
detectors is corrupted by stochastic noise. For example, let $\mathcal
{E}$ be the true energy of an electron hitting the CMS ECAL. Then the
observed value of the energy follows to a good approximation the
Gaussian distribution $\mathcal{N}(\mathcal{E},\sigma^2(\mathcal{E}))$,
where the variance satisfies [\citet{CMS2008}]
%
%
\begin{equation}
\biggl( \frac{\sigma(\mathcal{E})}{\mathcal{E}} \biggr)^2 =
\biggl( \frac
{S}{\sqrt{\mathcal{E}}}
\biggr)^2 + \biggl( \frac{N}{\mathcal{E}} \biggr)^2 +
C^2, \label{eq:ECAL_res}
\end{equation}
where $S$, $N$ and $C$ are fixed constants. The measurement noise is
not always additive. Furthermore, for more sophisticated measurements,
such as the ones combining information from several subdetectors or
more than one particle, the distribution of the response is usually not
available in a closed form. Indeed, most analyses rely on detector
simulations or auxiliary measurements to determine the detector response.

It should be pointed out that not all LHC physics analyses directly
rely on unfolding. The common factor between the examples given in
Section~\ref{sec:intro} is that these are \emph{measurement\/} analyses
and not \emph{discovery\/} analyses, meaning that these are analyses
studying in detail the properties of some already known phenomenon. In
such a case, the experimental interest often lies in the detailed
particle-level shape of some distribution which can be obtained using
unfolding, while discovery analyses are almost exclusively carried out
in the smeared space. Unfolding nevertheless plays an indirect role in
attempts to discover new physics at the LHC. Namely, discovery analyses
often use unfolded results as inputs to their analysis chain.

The need to unfold the measurements usually arises for the purposes of
the following:
\begin{itemize}
\item\textit{Comparison of experiments with different responses:} The
only direct way of comparing the spectra measured in two different
experiments is to compare the unfolded measurements.
\item\textit{Input to a subsequent analysis:} Certain tasks, such as
the estimation of parton distributions functions or the fine-tuning of
Monte Carlo event generators, typically require unfolded input spectra.
\item\textit{Comparison with future theories:} When unfolded spectra
are published, theorists can directly use them to compare with any new
theoretical predictions which might not have existed at the time of the
original measurement. This justification is sometimes considered
controversial since, alternatively, one could publish the response of
the detector and the theorists could use it to smear their new predictions.
\item\textit{Exploratory data analysis:} The unfolded spectrum could
reveal hidden features and structure in the data which are not
considered in any of the existing theoretical predictions.
\end{itemize}

According to the CERN Document Server (\url{https://cds.cern.ch/}), the
CMS experiment published in 2012 a total of 103 papers out of which 16
made direct use of unfolding and many more indirectly relied on
unfolded results. That year, unfolding was most often used in studies
of quantum chromodynamics (4~papers), forward physics (4) and
properties of the top quark (3). Most of these results relied on the
questionable bin-by-bin heuristic (8), while the EM algorithm (3) and
various forms of penalization (6) were also used. We expect similar
statistics to also hold for the other LHC experiments.

\section{Problem formulation} \label{sec:formulation}

In most situations in high energy physics, the data generation
mechanism can be modeled as a \emph{Poisson point process} [see, e.g.,
\citet{Reiss1993}]. Let $E$ be a compact interval on $\mathbb{R}$, $f$
a nonnegative function in $L^2(E)$ and $M$ a discrete random measure
on $E$. Then $M$ is a Poisson point process on state space $E$ with
intensity function $f$ if and only~if:
\begin{enumerate}
\item$M(B) \sim\operatorname{Poisson}(\lambda(B))$ with $\lambda
(B) =
{\int_{B}^{} f(s) \,\mathrm{d}s}$ for every Borel set $B \subset E$;
\item$M(B_1),\ldots,M(B_n)$ are independent for pairwise disjoint
Borel sets $B_i \subset E, i=1,\ldots,n$.
\end{enumerate}
In other words, the number of points $M(B)$ observed in the set $B
\subset E$ is Poisson distributed with mean ${\int_{B}^{} f(s)\, \mathrm
{d}s}$ and
the number of points in disjoint sets are independent random variables.

For the problem at hand, the Poisson process $M$ represents the true,
particle-level spectrum of events. The smeared, detector-level spectrum
is represented by another Poisson process $N$. The process $N$ is
assumed to have a state space $F$, which is a compact interval on
$\mathbb{R}$, and a nonnegative intensity function $g \in L^2(F)$. The
intensities of the two processes are related by a bounded linear
operator $K:L^2(E) \rightarrow L^2(F)$ so that $g = Kf$. In what
follows, we assume $K$ to be a Fredholm integral operator, that is,
%
%
\begin{equation}
g(t) = (Kf) (t) = {\int_{E}^{} k(t,s)f(s) \,\mathrm{d}s}, \label{eq:FredIntEq}
\end{equation}
where the kernel $k \in L^2(F \times E)$. For the purposes of this
paper, we assume that $k$ is known, although in reality there is
usually an uncertainty associated with it; see Section~\ref
{sec:discConc}. The unfolding problem is then to estimate the true
intensity $f$ given a single observation of the smeared Poisson process~$N$.

This Poisson inverse problem [\citet{Antoniadis2006,Reiss1993}] is
ill-posed in the sense that in virtually all practical cases the
pseudoinverse $K^{\dagger}$ of the forward operator $K$ is an unbounded,
and hence discontinuous, linear operator [\citet{Engl2000}]. This means
that the na\"ive approach of first estimating $g$ using, for example, a
kernel density estimate $\hat{g}$ and then estimating $f$ using $\hat
{f} = K^{\dagger}\hat{g}$ is unstable with respect to fluctuations of
$\hat{g}$.

To better understand the physical meaning of the kernel $k$, let us
consider the unfolding problem at the point level. Denoting by $X_i$
the true observables, the Poisson point process $M$ can be written as
$M = \sum_{i=1}^\tau\delta_{X_i}$, where $\delta_{X_i}$ is the Dirac
measure at $X_i \in E$ and $\tau,X_1,X_2,\ldots$ are independent random
variables such that $\tau\sim\operatorname{Poisson}(\lambda(E))$
and the
$X_i$ are identically distributed with probability density $f(\cdot
)/\lambda(E)$, where $\lambda(E) = {\int_{E}^{} f(s) \,\mathrm{d}s}$.

When the particle corresponding to $X_i$ traverses the detector, the
first thing that can happen is that it might not be observed at all due
to the limited efficiency and acceptance of the device. Mathematically,
this corresponds to \emph{thinning} of the Poisson process. Let $Z_i
\in
\{0,1\}$ be an indicator variable showing whether the point $X_i$ is
observed ($Z_i = 1$) or not ($Z_i = 0$). We assume that $\tau
,(X_1,Z_1),(X_2,Z_2),\ldots$ are independent and that the pairs
$(X_i,Z_i)$ are identically distributed. Then the thinned true process
is given by $M^* = \sum_{i=1}^\tau Z_i \delta_{X_i} = \sum_{i=1}^\xi
\delta_{X_i^*}$, where $\xi= \sum_{i=1}^\tau Z_i$ and the $X_i^*$ are
the true points with $Z_i = 1$. The thinned process $M^*$ is a Poisson
point process with intensity function $f^*(s) = \varepsilon(s) f(s)$,
where $\varepsilon(s) = P(Z_i = 1 | X_i = s)$ is the efficiency of the
detector for a true observation at $s \in E$.

For each observed point $X_i^* \in E$, the detector measures a noisy
value $Y_i \in F$. We assume that the smeared observations
$Y_i$ are i.i.d. with probability density
%
%
\begin{equation}
p(Y_i = t) = {\int_{E}^{} p\bigl(Y_i = t| X_i^* = s\bigr) p\bigl
(X_i^* = s\bigr) \,\mathrm{d}s}.
\end{equation}
From this, it follows that the smeared observations $Y_i$ constitute a
Poisson point process $N = \sum_{i=1}^\xi\delta_{Y_i}$ whose intensity
function $g$ is given by
%
%
\begin{equation}
g(t) = {\int_{E}^{} p\bigl(Y_i=t|X_i^*=s\bigr) \varepsilon(s)f(s)
\,\mathrm{d}s}.
\end{equation}
We hence identify that the kernel $k$ in equation \eqref{eq:FredIntEq}
is given by
%
%
\begin{equation}
k(t,s) = p\bigl(Y_i=t|X_i^*=s\bigr)\varepsilon(s).
\end{equation}
Notice that in the special case where $k(t,s) = k(t-s)$, unfolding
becomes a deconvolution problem [\citet{Meister2009}] for Poisson point
process observations.

\section{Unfolding methodology} \label{sec:EBU}

\subsection{Outline of the proposed methodology} \label{sec:outline}

In this section we propose a novel combination of statistical methods
for solving the high energy physics unfolding problem formalized in
Section~\ref{sec:formulation}. The proposed methodology is based on the
following key ingredients:
\begin{enumerate}
\item Discretization:
\begin{enumerate}[(a)]
\item[(a)] The smeared observations are discretized using a histogram.
\item[(b)] The unknown particle-level intensity is modeled using a B-spline,
that is, $f(s) = \sum_{j=1}^p \beta_j B_j(s), s \in E$, where $B_j(s),
j=1,\ldots,p$, are the B-spline basis functions.
\end{enumerate}
\item Point estimation:
\begin{enumerate}
\item[(a)] Posterior mean estimation of the unknown basis coefficients
$\bolds{\beta} = [ \beta_1, \ldots,   \beta_p ] ^{\mathrm{T}}$
using a single-component Metropolis--Hastings sampler.
\item[(b)] Empirical Bayes estimation of the scale $\delta$ of the
regularizing smoothness prior $p(\bolds{\beta}|\delta)$ using a Monte
Carlo EM algorithm.
\end{enumerate}
\item Uncertainty quantification:
\begin{enumerate}
\item[(a)] Iterative bias-correction of the point estimate $\hat{\bolds
{\beta}}$.
\item[(b)] Use of bootstrap percentile intervals of the bias-corrected
intensity function to form pointwise frequentist confidence bands for~$f$.
\end{enumerate}
\end{enumerate}

This methodology enables a principled solution of the unfolding
problem, including the choice of the regularization strength and
frequentist uncertainty quantification. We explain below each of these
steps in detail and argue why this particular choice of techniques
provides a natural framework for solving the problem.

\subsection{Discretization of the problem} \label{sec:discretization}

In applied situations, Poisson inverse problems are almost exclusively
studied in a form where both the observable process $N$ and the
unobservable process $M$ are discretized. Usually the first step is to
discretize the observable process using a histogram. In most
applications, this has to be done due to the discrete nature of the
detector. In our case, the observations are, at least in principle,
continuous, but we still carry out the discretization due to
computational reasons. Indeed, in many analyses, there can be millions
of observed collision events and treating each of these individually
would not be computationally feasible.

In order to discretize the smeared process $N$, let $\{F_i\}_{i=1}^n$
be a partition of the smeared space $F$ into $n$ ordered intervals and
let $y_i$ denote the number of points falling on interval $F_i$, that
is, $y_i = N(F_i), i=1,\ldots,n$. This can be seen as recording the
observed points in a histogram with bin contents $\mathbf{y} =
{[y_1,\ldots,y_n ] ^{\mathrm{T}}}$ and is indeed the form of
discretization most often employed in HEP.
This discretization is convenient since it now follows from $N$ being a
Poisson process that the $y_i$ are independent and Poisson distributed
with means
%
%
\begin{equation}
\mu_i = {\int_{F_i}^{} g(t) \,\mathrm{d}t} =
{\int_{F_i}^{} {\int_{E}^{} k(t,s)f(s) \,\mathrm{d}s} \,\mathrm{d}t},\qquad i
= 1,
\ldots,n. \label{eq:smearedMean}
\end{equation}

In the true space $E$, there is no need to settle only for histograms.
Instead, we consider a basis expansion of the true intensity $f$, that
is, $f(s) = \sum_{j=1}^p \beta_j \phi_j(s), s \in E$, where $\{\phi
_j\}
_{i=1}^p$ is a sufficiently large dictionary of basis functions.
Substituting the basis expansion into equation~\eqref{eq:smearedMean},
we find that the means $\mu_i$ are given by
%
%
\begin{equation}
\mu_i = \sum_{j=1}^p \biggl(
{\int_{F_i}^{} {\int_{E}^{} k(t,s)\phi_j(s) \,\mathrm{d}s}\, \mathrm
{d}t} \biggr) \beta_j = \sum_{j=1}^p
K_{i,j} \beta_j,
\end{equation}
where we have denoted
%
%
\begin{equation}
K_{i,j} = {\int_{F_i}^{} {\int_{E}^{} k(t,s) \phi_j(s) \,\mathrm{d}s}
\,\mathrm{d}t},\qquad i=1,\ldots,n, j=1,\ldots,p. \label{eq:Kij}
\end{equation}
Consequently, unfolding reduces to estimating $\bolds{\beta}$ in the
Poisson regression problem
%
%
\begin{equation}\label{eq:PoissonRegressionProblem}
\mathbf{y}|\bolds{\beta} \sim\operatorname{Poisson}(\mathbf
{K}\bolds{\beta})\vadjust{\goodbreak}
\end{equation}
for an ill-conditioned matrix $\mathbf{K} = (K_{i,j})$.

Since spectra in high energy physics are typically smooth functions,
splines [\citet{deBoor2001,Schumaker2007,Wahba1990}] provide a
particularly attractive way of representing the unknown intensity $f$.
Let $\min E = s_0 < s_1 < s_2 < \cdots< s_L < s_{L+1} = \max E$ be a
sequence of knots in the true space $E$. Then an order-$m$ spline with
knots $s_i, i=0,\ldots,L+1$, is a piecewise polynomial whose
restriction to each interval $[s_i,s_{i+1}), i=0,\ldots,L$, is an
order-$m$ polynomial (i.e., a polynomial of degree $m-1$) and which has
$m-2$ continuous derivatives at each interior knot $s_i, i=1,\ldots,L$.
An order-$m$ spline with $L$ interior knots has $p=L+m$ degrees of
freedom. In this work, we use exclusively order-4 cubic splines which
consist of third degree polynomials and are twice continuously
differentiable. Note also that an order-1 spline yields a histogram
representation of~$f$.

There exist various bases $\{\phi_j\}_{j=1}^p$ for expressing splines
of arbitrary order. We use B-splines $B_j, j=1,\ldots,p$, that is,
spline basis functions of minimal local support, because of their good
numerical properties and conceptual simplicity. \citeauthor{OSullivan1986}
(\citeyear
{OSullivan1986,OSullivan1988}) was among the first authors to use
regularized B-spline estimators in statistical applications, with the
approach later popularized by \citet{Eilers1996}. In the HEP unfolding
literature, penalized maximum likelihood estimation with B-splines goes
back to the work of \citet{Blobel1985} and recent contributions using
similar methodology include \citet{Dembinski2011} and \citet
{Milke2013}. We use the {\sc Matlab} Curve Fitting Toolbox to
efficiently evaluate and perform basic operations on B-splines. These
algorithms rely on the recursive use of lower-order B-spline basis
functions; for details, see \citet{deBoor2001}.

The nonnegativity of the intensity function $f$ is enforced by
constraining $\bolds{\beta}$ to be in $\mathbb{R}_+^p = \{ x \in
\mathbb
{R}^p: x_i \geq0, i=1,\ldots,p \}$. Since each of the
B-spline basis functions $B_j, j=1,\ldots,p$, is nonnegative, this is
a sufficient condition for the nonnegativity of $f$. It should be
noted, however, that generally this is not a necessary condition for
the nonnegativity of $f$ (except for order-1 and order-2 B-splines).
That is, when imposing $\bolds{\beta} \in\mathbb{R}_+^p$, we are
restricting ourselves to a proper subset of the set of positive splines
which may incur a slight, but not restrictive, reduction in the
versatility of the family of functions available to us [\citet{deBoor1974}].

\subsection{Point estimation}

\subsubsection{Posterior mean estimation of the spline coefficients}
\label{sec:estimation}

In contrast to most work on unfolding, we take a Bayesian approach to
estimation of the spline coefficients $\bolds{\beta}$. That is, we
estimate $\bolds{\beta}$ using the Bayesian\vadjust{\goodbreak} posterior\looseness=1
%
%
\begin{equation} \label{eq:BayesRule}
p(\bolds{\beta}|\mathbf{y},\delta) = \frac{p(\mathbf{y}|\bolds
{\beta})p(\bolds{\beta
}|\delta)}{p(\mathbf{y}|\delta)} = \frac{p(\mathbf{y}|\bolds
{\beta})p(\bolds{\beta
}|\delta)}{{\int_{\mathbb{R}_+^p}^{} p(\mathbf{y}|\bolds{\beta
}')p(\bolds{\beta }'|\delta) \,\mathrm{d}\bolds{\beta}'}},\qquad
\bolds{\beta} \in\mathbb{R}_+^p,
\end{equation}\looseness=0
where the likelihood is given by the Poisson regression model \eqref
{eq:PoissonRegressionProblem},
%
%
\begin{equation}
p(\mathbf{y}|\bolds{\beta}) = \prod_{i=1}^n
\frac{( \sum_{j=1}^p K_{i,j}\beta
_j )^{y_i}}{y_i!} e^{-\sum_{j=1}^p K_{i,j} \beta_j}, \qquad\bolds{\beta}
\in\mathbb{R}_+^p.
\end{equation}
The prior $p(\bolds{\beta}|\delta)$, which regularizes the otherwise
ill-posed problem, depends on a hyperparameter $\delta$, which controls
the concentration of the prior and is analogous to the regularization
parameter in the classical methods for solving inverse problems.

We decided to use the Bayesian approach for two reasons. First, it
provides a natural interpretation for the regularization via the prior
density $p(\bolds{\beta}|\delta)$, which should be chosen in such a way
that most of its probability mass lies in physically plausible regions
of the parameter space $\mathbb{R}_+^p$. Second, the Bayesian framework
enables a straightforward, data-driven way of choosing the
regularization strength $\delta$ using empirical Bayes estimation as
explained below in Section~\ref{sec:empiricalBayes}.

In order to regularize the problem, let $\bolds{\beta} \in\mathbb
{R}_+^p$, and $\delta> 0$, and consider the truncated Gaussian
smoothness prior
%
%
\begin{equation}
\quad p(\bolds{\beta}|\delta) \propto\exp\bigl( -\delta\bigl\|f''
\bigr\|_2^2 \bigr) = \exp\bigl( -\delta{\int_{E}^{} \bigl \{f''(s)
\bigr\}^2 \,\mathrm{d}s} \bigr) = \exp\bigl( -
\delta{\bolds{\beta}^{\mathrm{T}}} \bolds{\Omega} \bolds{\beta}
\bigr),\label
{eq:smoothnessPrior}
\end{equation}
where the elements of the $p \times p$ matrix $\bolds{\Omega}$ are given
by $\Omega_{i,j} = {\int_{E}^{} B_i''(s) B_j''(s) \,\mathrm{d}s}$. The
interpretation of this prior is that the total curvature of $f$,
characterized by $\|f''\|_2^2$, should be small. In other words, $f$
should be a relatively smooth function. The strength of the
regularization is controlled by the hyperparameter $\delta$---the
larger the value of $\delta$, the smoother $f$ is required to~be.

The prior as defined by equation \eqref{eq:smoothnessPrior} does not
enforce any boundary conditions for the unknown intensity $f$. As a
result, the matrix $\bolds{\Omega}$ has rank $p-2$, and hence the
prior is
potentially improper (this depends on the orientation of the null space
of $\bolds{\Omega}$). Although the posterior would still be a proper
probability density, the rank deficiency of $\bolds{\Omega}$ is
undesirable since the empirical Bayes approach requires a proper prior
distribution. Furthermore, without any boundary constraints, the
unfolded intensity has an unnecessarily large variance near the boundaries.

To address these issues, we use \emph{Aristotelian boundary conditions}
[\citet{Calvetti2006}], where the idea is to condition the smoothness
penalty on the boundary values $f(s_0)$ and $f(s_{L+1})$ and then
introduce additional hyperpriors for these values. Since $f(s_0) =
\beta
_1 B_1(s_0)$ and $f(s_{L+1}) = \beta_p B_p(s_{L+1})$, we can
equivalently condition on $(\beta_1,\beta_p)$. As a result, the prior
model becomes
%
%
\begin{equation}
p(\bolds{\beta}|\delta) = p(\beta_2,\ldots,\beta_{p-1}|
\beta_1,\beta_p,\delta)p(\beta_1|\delta)p(
\beta_p|\delta), \qquad\bolds{\beta} \in\mathbb{R}_+^p,
\end{equation}
where $p(\beta_2,\ldots,\beta_{p-1}|\beta_1,\beta_p,\delta)
\propto
\exp( -\delta{\bolds{\beta}^{\mathrm{T}}} \bolds{\Omega} \bolds
{\beta}
)$. We
model the boundaries using once again truncated Gaussians:
%
%
\begin{eqnarray}
p(\beta_1|\delta) &\propto&\exp\bigl(-\delta\gamma_\mathrm{L}
\beta_1^2 \bigr),\qquad \beta_1 \geq0,
\\
p(\beta_p|\delta) &\propto&\exp\bigl(-\delta\gamma_\mathrm{R}
\beta_p^2 \bigr), \qquad\beta_p \geq0,
\end{eqnarray}
where $\gamma_\mathrm{L}, \gamma_\mathrm{R} > 0$ are fixed constants.
The full prior can then be written as
%
%
\begin{equation}
p(\bolds{\beta}|\delta) \propto\exp\bigl( -\delta{\bolds{\beta
}^{\mathrm{T}}}
\bolds{
\Omega}_\mathrm{A} \bolds{\beta} \bigr),\qquad \bolds{\beta} \in\mathbb{R}_+^p,
\label{eq:AristotelianPrior}
\end{equation}
where the elements of the $p \times p$ matrix $\bolds{\Omega}_\mathrm{A}$
are given by
%
%
\begin{equation}
\Omega_{\mathrm{A},i,j} = %
\cases{ \Omega_{i,j} +
\gamma_\mathrm{L}, &\quad$\mbox{if } i=j=1$, \vspace*{2pt}
\cr
\Omega_{i,j} + \gamma_\mathrm{R}, &\quad$\mbox{if } i=j=p$,
\vspace*{2pt}
\cr
\Omega_{i,j}, &\quad$\mbox{otherwise}$.} %
\end{equation}
The augmented matrix $\bolds{\Omega}_\mathrm{A}$ is positive
definite, and
hence equation \eqref{eq:AristotelianPrior} defines a proper
probability density.

Once the hyperparameter $\delta$ has been estimated using empirical
Bayes (see Section~\ref{sec:empiricalBayes}), we plug its estimate
$\hat
{\delta}$ into Bayes' rule \eqref{eq:BayesRule} to obtain the empirical
Bayes posterior $p(\bolds{\beta}|\mathbf{y},\hat{\delta})$. We
then use the
mean of this posterior as a point estimator $\hat{\bolds{\beta}}$ of the
spline coefficients $\bolds{\beta}$, that is, $\hat{\bolds{\beta}}
= \E
(\bolds{\beta}|\mathbf{y},\hat{\delta} )$, yielding the estimator
$\hat
{f}(s) = \sum_{j=1}^p \hat{\beta}_j B_j(s)$ of the unknown intensity $f$.

Of course, in practice, the posterior $p(\bolds{\beta}|\mathbf
{y},\delta)$ is
not available in a closed form because of the intractable integral in
the denominator of Bayes' rule~\eqref{eq:BayesRule}. Hence, we need to
resort to Markov chain Monte Carlo (MCMC) [\citet{Robert2004}] sampling
from the posterior and the posterior mean is then computed as the
empirical mean of the Monte Carlo sample. Unfortunately, the most
elementary MCMC samplers are not well-suited for solving the unfolding
problem: Gibbs sampling is not computationally tractable since the full
posterior conditionals do not belong to any of the standard families of
probability distributions and the Metropolis--Hastings sampler with
multivariate proposals is difficult to implement since the posterior
can have very different scales for different components of $\bolds
{\beta}$.

To be able to efficiently sample from the posterior, we adopt the
single-component Metropolis--Hastings sampler (also known as the
Metropolis-within-Gibbs sampler) proposed by \citet{Saquib1998}.
Denoting $\bolds{\beta}_{-k} = {[\beta_1, \ldots, \beta_{k-1},
\beta_{k+1}, \ldots, \beta _p ] ^{\mathrm{T}}}$, the basic idea of
the sampler is to approximate the full posterior
conditionals $p(\beta_k|\bolds{\beta}_{-k},\mathbf{y},\delta)$ of
the Gibbs
sampler using a more tractable density [\citet
{Gilks1996MCMC,Gilks1996FullCondd}]. One then samples from this
approximate full conditional and performs a Metropolis--Hastings
acceptance step to correct for the approximation error. In our case, we
take a second-order Taylor expansion of the log full conditional,
resulting in a Gaussian approximation of this density. When the mean of
the Gaussian is nonnegative, we sample from its truncation to the
nonnegative real line, and if the mean is negative, we replace the
Gaussian tail by an exponential distribution. Further details on the
MCMC sampler can be found in Section~III.C of \citet{Saquib1998}, while
the online supplement [\citet{Kuusela2015}] provides details on the
convergence and mixing checks that were performed for the sampler.

\subsubsection{Empirical Bayes selection of the regularization
strength} \label{sec:empiricalBayes}

The\break Bayesian approach to solving inverse problems is
particularly attractive since it admits selection of the regularization
strength $\delta$ using marginal maximum likelihood estimation. For a
comprehensive introduction to this and related empirical Bayes methods,
see, for example, Chapter~5 of \citet{Carlin2009}. The main idea in
empirical Bayes is to regard the marginal distribution $p(\mathbf
{y}|\delta
)$ appearing in the denominator of Bayes' rule \eqref{eq:BayesRule} as
a parametric model for the data $\mathbf{y}$ and then use standard
frequentist point estimation techniques to estimate the hyperparameter
$\delta$.

The \emph{marginal maximum likelihood estimator} (MMLE) of the
hyperparameter $\delta$ is defined as the maximizer of $p(\mathbf
{y}|\delta
)$ with respect to $\delta$. That is, we estimate $\delta$ using
%
%
\begin{equation}
\hat{\delta} = \argmax_{\delta> 0} p(\mathbf{y}|\delta) =
\argmax_{\delta
> 0} {\int_{\mathbb{R}_+^p}^{} p(\mathbf{y}| \bolds{\beta
})p(\bolds{\beta}|\delta) \,\mathrm{d}\bolds{\beta}}. \label
{eq:defMMLE}
\end{equation}
Computing the MMLE is nontrivial since we cannot evaluate the
high-dimensional integral in \eqref{eq:defMMLE} either in a closed form
or using standard numerical integration methods. Monte Carlo
integration, where one samples $\{\bolds{\beta}^{(s)}\}_{s=1}^{S}$ from
the prior $p(\bolds{\beta}|\delta)$ and then approximates
%
%
\begin{equation}
p(\mathbf{y}|\delta) \approx\frac{1}{S} \sum_{s=1}^{S}
p\bigl(\mathbf{y}|\bolds{\beta}^{(s)}\bigr),\qquad \bolds{
\beta}^{(s)} \overset{\mathrm{i.i.d.}} {\sim} p(\bolds{\beta
}|\delta),
\label{eq:naiveMCIntegration}
\end{equation}
is also out of question. This is because, in the high-dimensional
parameter space, most of the $\bolds{\beta}^{(s)}$'s fall on regions where
the likelihood $p(\mathbf{y}|\bolds{\beta}^{(s)})$ is numerically
zero. Hence,
we would need an enormous sample size $S$ to get even a rough idea of
the marginal likelihood $p(\mathbf{y}|\delta)$.

Luckily, it is possible to circumvent these issues by using the
expectation--maximization (EM) algorithm [\citet
{Dempster1977,McLachlan2008}] to find the MMLE. In the context of
Poisson inverse problems, this approach was originally proposed by
\citeauthor{Geman1985} (\citeyear{Geman1985,Geman1987}) for tomographic image reconstruction and
later studied and extended by \citet{Saquib1998}, but has received
little attention since then. When applied to the unfolding problem, the
standard EM prescription reads as follows. Let $(\mathbf{y},\bolds
{\beta})$ be
the complete data, in which case the complete-data log-likelihood is
given by
%
%
\begin{equation}
l(\delta; \mathbf{y},\bolds{\beta}) = \log p(\mathbf{y},\bolds
{\beta}|\delta) =
\log p(\mathbf{y}|\bolds{\beta}) + \log p(\bolds{\beta}|\delta),
\end{equation}
where we have used $p(\mathbf{y},\bolds{\beta}|\delta) = p(\mathbf
{y}|\bolds{\beta
})p(\bolds{\beta}|\delta)$. In the E-step of the algorithm, one computes
the expectation of the complete-data log-likelihood over the unknown
spline coefficients $\bolds{\beta}$ conditional on the observations
$\mathbf
{y}$ and the current hyperparameter $\delta^{(i)}$:
%
%
\begin{eqnarray}
Q\bigl(\delta;\delta^{(i)}\bigr) = \E\bigl(l(\delta; \mathbf{y},
\bolds{\beta}) | \mathbf{y},\delta^{(i)} \bigr) &= &\E\bigl(\log p(
\mathbf{y},\bolds{\beta}|\delta) | \mathbf{y},\delta^{(i)} \bigr)
\\
&=& \E\bigl(\log p(\bolds{\beta}|\delta) | \mathbf{y},\delta^{(i)}
\bigr) + \mathrm{const},
\end{eqnarray}
where the constant does not depend on $\delta$. In the subsequent
M-step, one maximizes the expected complete-data log-likelihood
$Q(\delta;\delta^{(i)})$ with respect to the hyperparameter $\delta$.
This maximizer is then used as the hyperparameter estimate on the next
step of the algorithm:
%
%
\begin{equation}
\delta^{(i+1)} = \argmax_{\delta> 0} Q\bigl(\delta;
\delta^{(i)}\bigr) = \argmax_{\delta> 0} \E\bigl(\log p(\bolds{
\beta}|\delta) | \mathbf{y},\delta^{(i)} \bigr). \label{eq:MStep}
\end{equation}
By Theorem~1 of \citet{Dempster1977}, each step of this iteration is
guaranteed to increase the incomplete-data likelihood $p(\mathbf
{y}|\delta
)$, that is, $p(\mathbf{y}|\delta^{(i+1)}) \geq p(\mathbf{y}|\delta^{(i)}),
t=0,1,2,\ldots.$ With this construction, the incomplete-data likelihood
conveniently coincides with the marginal likelihood and, hence, the EM
algorithm enables us find the MMLE $\hat{\delta}$ of the hyperparameter~$\delta$.

The expectation in equation \eqref{eq:MStep},
%
%
\begin{equation}
\E\bigl(\log p(\bolds{\beta}|\delta) | \mathbf{y},\delta^{(i)}
\bigr)
= {\int_{\mathbb{R}_+^p}^{} p\bigl(\bolds{\beta}|\mathbf{y},
\delta^{(i)}\bigr)\log p(\bolds{\beta}|\delta) \,\mathrm{d}\bolds
{\beta}},
\label{eq:expectationIntegral}
\end{equation}
again involves an intractable integral, but this time can be computed
using Monte Carlo integration. We simply need to sample $\{\bolds
{\beta
}^{(s)}\}_{s=1}^S$ from the posterior $p(\bolds{\beta}|\mathbf
{y},\delta
^{(i)})$ and then replace the expectation by its Monte Carlo approximation:
%
%
\begin{equation}
\E\bigl(\log p(\bolds{\beta}|\delta) | \mathbf{y},\delta^{(i)}
\bigr)
\approx\frac{1}{S} \sum_{s=1}^S \log
p\bigl(\bolds{\beta}^{(s)}|\delta\bigr),\qquad \bolds{\beta}^{(s)}
\sim p\bigl(\bolds{\beta}|\mathbf{y},\delta^{(i)}\bigr). \label
{eq:MCIntegration}
\end{equation}
This Monte Carlo approximation is better behaved than that of
equation~\eqref{eq:naiveMCIntegration} due to the appearance of the
logarithm and due to the fact that the sampling is from the posterior
instead of the prior. The posterior sample can be obtained using the
single-component Metropolis--Hastings sampler described in Section~\ref
{sec:estimation}. The resulting variant of the EM~algorithm is called a
\emph{Monte Carlo expectation--maximization \emph{(MCEM)} algorithm}
[\citet
{Wei1990}].

To summarize, the MCEM algorithm for finding the MMLE of the
hyperparameter $\delta$ iterates between the following two steps:
\begin{longlist}[M-step:]
\item[\textit{E-step}:]Sample $\bolds{\beta}^{(1)},\ldots,\bolds{\beta
}^{(S)}$ from
the posterior $p(\bolds{\beta}|\mathbf{y},\delta^{(i)})$ and compute
%
%
\begin{equation}
\widetilde{Q}\bigl(\delta;\delta^{(i)}\bigr) = \frac{1}{S} \sum
_{s=1}^S \log p\bigl(\bolds{
\beta}^{(s)}|\delta\bigr). \label{eq:QTilde}
\end{equation}
\item[\textit{M-step}:] Set $\delta^{(i+1)} = \argmax_{\delta> 0} \widetilde
{Q}(\delta;\delta^{(i)})$.
\end{longlist}
This algorithm has a rather intuitive interpretation. First, on the
E-step, we use the current iterate $\delta^{(i)}$ to produce a sample
of $\bolds{\beta}$'s from the posterior. Since this sample summarizes our
current best understanding of $\bolds{\beta}$, we then tune the prior by
varying $\delta$ on the M-step to match this sample as well as
possible, and the value of $\delta$ that matches the posterior sample
the best will then become the next iterate~$\delta^{(i+1)}$.

When $p(\bolds{\beta}|\delta)$ is given by the Aristotelian smoothness
prior \eqref{eq:AristotelianPrior}, the M-step of the MCEM
algorithm is available in a closed form. Taking normalization into
account, the prior density is given by
%
%
\begin{equation}
p(\bolds{\beta}|\delta) = C(\delta) \exp\bigl(-\delta{\bolds {\beta
}^{\mathrm{T}}} \bolds{
\Omega}_\mathrm{A} \bolds{\beta}\bigr),
\end{equation}
where the normalization constant $C(\delta)$ depends on the
hyperparameter $\delta$ and satisfies $C(\delta) = \delta^{p/2} /
{\int_{\mathbb{R}_+^p}^{} \exp(-{\bolds{\beta}^{\mathrm
{T}}}\bolds{\Omega }_\mathrm{A}\bolds {\beta}) \,\mathrm{d}\bolds
{\beta}}$. Hence,
%
%
\begin{equation}
\log p(\bolds{\beta}|\delta) = \frac{p}{2} \log\delta- \delta
{\bolds{ \beta}^{\mathrm{T}}} \bolds{\Omega}_\mathrm{A} \bolds
{\beta} + \mathrm{const},
\end{equation}
where the constant does not depend on $\delta$. Plugging this into
equation \eqref{eq:QTilde}, we find that the maximizer on the M-step is
given by
%
%
\begin{equation}
\delta^{(i+1)} = \frac{1}{({2}/{(pS)}) \sum_{s=1}^S {(\bolds
{\beta }^{(s)})^{\mathrm{T}}} \bolds{\Omega}_\mathrm{A} \bolds
{\beta}^{(s)}}.
\end{equation}

\begin{algorithm}[t]
\caption{MCEM algorithm for finding the MMLE}
\label{alg:MCEM}
\begin{algorithmic}
\Require
\State$\mathbf{y}$---Observed data;
\State$\delta^{(0)} > 0$---Initial guess;
\State$N_\mathrm{EM}$---Number of MCEM iterations;
\State$S$---Size of the MCMC sample;
\State$\bolds{\beta}_\mathrm{init}$---Starting point for the MCMC sampler;
\Ensure
\State$\hat{\delta}$---MMLE of the hyperparameter $\delta$;
\State
\State Set $\bar{\bolds{\beta}} = \bolds{\beta}_\mathrm{init}$;
\For{$i = 0$ \algorithmicto\ $N_\mathrm{EM}-1$}
\State Sample $\bolds{\beta}^{(1)},\bolds{\beta}^{(2)},\ldots
,\bolds{\beta
}^{(S)} \sim p(\bolds{\beta}|\mathbf{y},\delta^{(i)})$ starting
from $\bar{\bolds
{\beta}}$ using the single-component Metropolis--Hastings sampler of
\citet{Saquib1998};
\State Set
\[
\delta^{(i+1)} = \frac{1}{({2}/{(pS)}) \sum_{s=1}^S {(\bolds
{\beta }^{(s)})^{\mathrm{T}}} \bolds{\Omega}_\mathrm{A} \bolds
{\beta}^{(s)}};
\]
\State Compute $\bar{\bolds{\beta}} = \frac{1}{S} \sum_{s=1}^S
\bolds{\beta}^{(s)}$;
\EndFor
\State\Return$\hat{\delta} = \delta^{(N_\mathrm{EM})}$.
\end{algorithmic}
\end{algorithm}

The resulting iteration for finding the MMLE $\hat{\delta}$ is
summarized in Algorithm~\ref{alg:MCEM}. The MCMC sampler is started
from the empirical mean of the posterior sample of the previous
iteration in order to facilitate the convergence of the Markov chain.
In this work, we run the MCEM algorithm for a fixed number of steps
$N_\mathrm{EM}$, but one could easily devise more elaborate stopping
rules for the algorithm. Note, however, that the optimal choice of this
stopping rule and the MCMC sample size $S$ are, to a large extent, open
problems [\citet{Booth1999}].

\subsection{Uncertainty quantification} \label{sec:UQ}

Frequentist uncertainty quantification in nonparametric inference is
generally considered to be a very challenging problem; see, for
example, Chapter~6 of \citet{Ruppert2003} for an overview of some of
the issues involved. Common approaches considered in the literature
include bootstrapping [\citet{Davison1997}] and various Bayesian
constructions [see, e.g., \citet{Wahba1983}, \citet{Wood2006} and
\citet
{Weir1997}]. In the case of classical nonparametric regression,
Bayesian intervals are often argued to have good frequentist properties
based on the results of \citet{Nychka1988}, which guarantee that the
coverage probability \emph{averaged over the design points} is close to
the nominal value. Such average coverage, however, provides no
guarantees about pointwise coverage and the intervals can suffer from
significant pointwise under- or overcoverage as demonstrated by \citet
{Ruppert2000}.

The main problem in building confidence intervals for the unknown
intensity $f$ is the bias that is present in the estimator $\hat{f}$ in
order to regularize the problem. We show using simulations in
Section~\ref{sec:simulations} that this bias is a major hurdle for both
Bayesian and bootstrap confidence intervals, resulting in major
frequentist undercoverage in regions of sizable bias. To overcome this
issue, we propose attacking the problem from a different perspective:
instead of directly using the variability of $\hat{f}$ to construct
confidence intervals, we first iteratively reduce the bias of $\hat{f}$
and then use the variability of the bias-corrected estimator $\hat
{f}_\mathrm{BC}$ to construct confidence intervals. This approach has
similarities with the recent work of \citet{Javanmard2014} on
uncertainty quantification in high-dimensional regression by de-biasing
the estimator, but our problem setting and bias-correction method are
different from theirs.

It may at first seem counterintuitive that reducing the bias of $\hat
{f}$ enables us to form improved confidence intervals---it is after
all the bias that regularizes the otherwise ill-posed problem. It is
indeed true that the iterative bias-correction described below
increases the variance of the point estimator $\hat{f}_\mathrm{BC}$;
but at the same time the coverage performance of the intervals formed
using $\hat{f}_\mathrm{BC}$ improves at each iteration of the
procedure. What is more, our simulations reported in Section~\ref
{sec:simulations} indicate that, by stopping the bias-correction
iteration early enough, it is possible to attain nearly nominal
coverage with only a modest increase in interval length. In other
words, one can use the iterative bias-correction to remove so much of
the bias that the interval coverage probability is close to its nominal
value, but the small amount of residual bias that remains in enough to
regularize the interval length. This phenomenon is consistent with what
\citet{Javanmard2014} observe in debiased $l^1$-regularized regression.

\subsubsection{Iterative bias-correction} \label{sec:bsBiasCorr}

Our bias-correction approach is based on an iterative use of the
bootstrap to estimate the bias of the point estimate $\hat{\bolds
{\beta
}}$. A similar approach has been previously used by \citet{Kuk1995} and
\citet{Goldstein1996} to debias estimators in generalized linear mixed
models, but, to the best of our knowledge, this procedure has not been
previously employed to improve the frequentist coverage of confidence
intervals in ill-posed nonparametric inverse problems, such as the one
studied here.

\begin{algorithm}[b]
\caption{Iterative bias-correction}
\label{alg:BsBC}
\begin{algorithmic}
\Require
\State$\hat{\bolds{\beta}}{}^{(0)}$---Observed value of the
estimator $\bolds
{\hat{\beta}}$;
\State$\hat{\delta}$---Estimated value of the hyperparameter
$\delta$;
\State$N_\mathrm{BC}$---Number of bias-correction iterations;
\State$R_\mathrm{BC}$---Size of the bootstrap sample;
\State$S$---Size of the MCMC sample;
\Ensure
\State$\hat{\bolds{\beta}}_\mathrm{BC}$---Bias-corrected point estimate;
\State
\For{$i = 0$ \algorithmicto\ $N_\mathrm{BC}-1$}
\State Sample $\mathbf{y}^{*(1)},\mathbf{y}^{*(2)},\ldots,\mathbf
{y}^{*(R_\mathrm
{BC})} \overset{\mathrm{i.i.d.}}{\sim} \operatorname
{Poisson}(\mathbf{K}\hat{\bolds
{\beta}}{}^{(i)})$;
\State For each $r=1,\ldots,R_\mathrm{BC}$, compute $\hat{\bolds
{\beta
}}{}^{*(r)} = \E(\bolds{\beta}|\mathbf{y}^{*(r)},\hat{\delta})$ by
sampling $S$
observations from the posterior using the single-component
Metropolis--Hastings sampler of \citet{Saquib1998};
\State Compute $\widehat{\mathrm{bias}}{}^{(i)}(\hat{\bolds{\beta}})
= \frac
{1}{R_\mathrm{BC}} \sum_{r=1}^{R_\mathrm{BC}} \hat{\bolds{\beta
}}{}^{*(r)} -
\hat{\bolds{\beta}}{}^{(i)}$;
\State Set $\hat{\bolds{\beta}}{}^{\prime(i+1)} = \hat{\bolds{\beta
}}{}^{(0)} -
\widehat{\mathrm{bias}}{}^{(i)}(\hat{\bolds{\beta}})$;
\State Set $\hat{\bolds{\beta}}{}^{(i+1)} = \mathbf{1} \{ \hat
{\bolds{\beta
}}{}^{\prime(i+1)} \geq\mathbf{0} \} \circ\hat{\bolds{\beta
}}{}^{\prime(i+1)}$,
where $\circ$ denotes the element-wise product;
\EndFor
\State\Return$\hat{\bolds{\beta}}_\mathrm{BC} = \hat{\bolds
{\beta
}}{}^{(N_\mathrm{BC})}$.
\end{algorithmic}
\end{algorithm}

By definition, the bias of $\hat{\bolds{\beta}}$ is given by
$\operatorname
{bias}(\hat{\bolds{\beta}}) = \E_{\bolds{\beta}} (\hat{\bolds
{\beta}}) - \bolds
{\beta}$. The standard bootstrap estimate of this bias [see, e.g.,
\citet{Davison1997}] replaces the actual value of $\bolds{\beta}$ by the
observed value of the estimator $\hat{\bolds{\beta}}$ which we
denote by
$\hat{\bolds{\beta}}{}^{(0)}$. In other words, the bootstrap estimate
of the
bias is
%
%
\begin{equation}
\widehat{\operatorname{bias}}{}^{(0)}(\hat{\bolds{\beta}}) = \E
_{\hat{\bolds{\beta
}}{}^{(0)}}
(\hat{\bolds{\beta}}) - \hat{\bolds{\beta}}{}^{(0)}, \label{eq:biasHat0}
\end{equation}
where in practice the expectation is usually replaced by its empirical
version obtained using simulations. The estimated bias can then be
subtracted from $\hat{\bolds{\beta}}{}^{(0)}$ to obtain a
bias-corrected estimator
%
%
\begin{equation}
\hat{\bolds{\beta}}{}^{(1)} = \hat{\bolds{\beta}}{}^{(0)} -
\widehat{\operatorname{bias}}{}^{(0)}(\hat{\bolds{\beta}}).
\end{equation}
Now, ignoring the bootstrap sampling error, the reason $\hat{\bolds
{\beta
}}{}^{(1)}$ is not a perfectly unbiased estimator of $\bolds{\beta}$ is the
fact that in equation \eqref{eq:biasHat0} the actual value of $\bolds
{\beta
}$ was replaced by an estimate $\hat{\bolds{\beta}}{}^{(0)}$. But
since $\bolds
{\hat{\beta}}{}^{(1)}$ should be better than $\hat{\bolds{\beta
}}{}^{(0)}$ at
estimating $\bolds{\beta}$ (at least in the sense that $\hat{\bolds
{\beta
}}{}^{(1)}$ should be less biased than~$\hat{\bolds{\beta}}{}^{(0)}$),
we can
replace $\hat{\bolds{\beta}}{}^{(0)}$ in equation \eqref{eq:biasHat0}
by $\bolds
{\hat{\beta}}{}^{(1)}$ to obtain a new estimate of the bias
%
%
\begin{equation}
\widehat{\operatorname{bias}}{}^{(1)}(\hat{\bolds{\beta}}) = \E
_{\hat{\bolds{\beta
}}{}^{(1)}}
(\hat{\bolds{\beta}}) - \hat{\bolds{\beta}}{}^{(1)}, \label{eq:biasHat1}
\end{equation}
which gives rise to a new bias-corrected estimator
%
%
\begin{equation}
\hat{\bolds{\beta}}{}^{(2)} = \hat{\bolds{\beta}}{}^{(0)} -
\widehat{\operatorname{bias}}{}^{(1)}(\hat{\bolds{\beta}}).
\end{equation}
The estimator $\hat{\bolds{\beta}}{}^{(2)}$ can then be used to
replace $\bolds
{\hat{\beta}}{}^{(1)}$ in equation \eqref{eq:biasHat1} which naturally
leads us to the following \emph{iterative bias-correction} procedure:
\begin{enumerate}
\item Estimate the bias: $\widehat{\operatorname{bias}}{}^{(i)}(\hat
{\bolds{\beta
}}) = \E_{\hat{\bolds{\beta}}{}^{(i)}} (\hat{\bolds{\beta}}) -
\hat{\bolds{\beta}}{}^{(i)}$.
\item Compute the bias-corrected estimate: $\hat{\bolds{\beta
}}{}^{(i+1)} =
\hat{\bolds{\beta}}{}^{(0)} - \widehat{\operatorname
{bias}}{}^{(i)}(\hat{\bolds{\beta}})$.
\end{enumerate}
The details of this procedure, when the expectation $\E_{\hat{\bolds
{\beta
}}{}^{(i)}} (\hat{\bolds{\beta}})$ is replaced by its empirical
version and
the positivity constraint of $\bolds{\beta}$ is enforced by setting any
negative entries to zero, are given in Algorithm \ref{alg:BsBC}. Note
that for computational reasons we keep the hyperparameter $\delta$
fixed to its estimated value $\hat{\delta}$ instead of resampling it.

\subsubsection{Pointwise confidence bands from the bias-corrected estimator}

The bias-corrected spline coefficients $\hat{\bolds{\beta}}_\mathrm{BC}$
are associated with a bias-corrected intensity function estimate $\hat
{f}_\mathrm{BC}(s) = \sum_{j=1}^p \hat{\beta}_{\mathrm{BC},j} B_j(s)$.
As the bias-corrected estimator $\hat{f}_\mathrm{BC}$ has a smaller
bias than the original estimator $\hat{f}$, the variability of $\hat
{f}_\mathrm{BC}$ can be used to construct confidence intervals for $f$
that have better coverage properties than the intervals based on the
variability of $\hat{f}$. The basic approach we follow is to use the
bootstrap to probe the sampling distribution of $\hat{f}_\mathrm{BC}$
in order to construct pointwise\footnote{For simplicity, we restrict
our attention to $1-2\alpha$ \emph{pointwise} frequentist confidence
bands, that is, collections of random intervals $[\b{\textit{f}}(s;
\mathbf
{y}), \bar{f}(s; \mathbf{y})]$ that for any $s \in E$, $\alpha\in(0,0.5)$
and intensity function~$f$ aim to satisfy $P_f(\b{\textit{f}}(s;
\mathbf
{y}) \leq f(s) \leq\bar{f}(s; \mathbf{y})) \geq1 - 2 \alpha$. This
is in
contrast with $1-2\alpha$ \emph{simultaneous} confidence bands, which
would satisfy for any $\alpha\in(0,0.5)$ and intensity function~$f$
the property $P_f(\b{\textit{f}}(s; \mathbf{y}) \leq f(s) \leq\bar{f}(s;
\mathbf{y}), \forall s \in E) \geq1 - 2 \alpha$; see also the discussion
in Section~\ref{sec:discConc}.} confidence bands for $f$.

To obtain the bootstrap sample, we generate $R_\mathrm{UQ}$ i.i.d.
observations from the $\operatorname{Poisson}(\hat{\bolds{\mu}})$
distribution,
where $\hat{\bolds{\mu}} = \mathbf{y}$ is the maximum likelihood
estimator of
$\bolds{\mu} = \mathbf{K}\bolds{\beta}$. For each resampled
observation $\mathbf
{y}^{*(r)}, r=1,\ldots,R_\mathrm{UQ}$, we compute a resampled point
estimate $\hat{\bolds{\beta}}{}^{*(r)} = \E(\bolds{\beta}|\mathbf
{y}^{*(r)},\hat
{\delta})$. Again, for computational reasons, the hyperparameter
$\delta
$ is kept fixed to its estimated value $\hat{\delta}$ for the observed
datum~$\mathbf{y}$. The bias-correction procedure described in Algorithm
\ref{alg:BsBC} is then run with the resampled point estimate $\hat
{\bolds
{\beta}}{}^{*(r)}$ to obtain a resampled bias-corrected point estimate
$\hat{\bolds{\beta}}{}^{*(r)}_\mathrm{BC}$, leading to a resampled
bias-corrected intensity function $\hat{f}{}^{*(r)}_\mathrm{BC}$.

The sample $\mathcal{F}_\mathrm{BC}^* = \{\hat{f}{}^{*(r)}_\mathrm
{BC}\}
_{r=1}^{R_{\mathrm{UQ}}}$ is a bootstrap representation of the sampling
distribution of~$\hat{f}_\mathrm{BC}$ and can be used to construct
various forms of approximate confidence intervals for $f$ [\citet
{Efron1993,Davison1997}]. The simplest approach is to simply use the
pointwise standard deviations of the bootstrap sample to form standard
error intervals for $f$. However, it was concluded using simulations
that the standard error intervals suffer from a slight reduction in
coverage due to the skewness of the sampling distribution induced by
the positivity constraint, especially in areas of low intensity values.
To account for this effect, we use instead the bootstrap percentile
intervals. More specifically, for every $s \in E$, an approximate
$1-2\alpha$ confidence interval for $f(s)$ is given by $[\hat
{f}_{\mathrm{BC},\alpha}^*(s),\hat{f}_{\mathrm{BC},1-\alpha}^*(s)]$,
where $\hat{f}_{\mathrm{BC},\alpha}^*(s)$ denotes the $\alpha$-quantile
of the bias-corrected bootstrap sample $\mathcal{F}_\mathrm{BC}^*$
evaluated at point~$s \in E$. The formal justification for these
intervals comes from an implicit use of a transformation that
(approximately) normalizes the sampling distribution of $\hat
{f}_\mathrm
{BC}(s)$; see Section~13.3 of \citet{Efron1993}.

\section{Simulation studies} \label{sec:simulations}

\subsection{Experiment setup} \label{sec:simSetup}

We first demonstrate the proposed unfolding\break methodology
using simulated data. The data were generated using a two-component
Gaussian mixture model on top of a uniform background and smeared by
convolving the particle-level intensity with a Gaussian density.
Specifically, the true process $M$ had the intensity
%
%
\begin{equation}
f(s) = \lambda_\mathrm{tot} \biggl\{ \pi_1
\mathcal{N}(s|{-2},1) + \pi_2 \mathcal{N}(s|2,1) + \pi_3
\frac{1}{|E|} \biggr\}, \qquad s \in E,
\end{equation}
where $\lambda_\mathrm{tot} = \E(\tau) = {\int_{E}^{} f(s) \,\mathrm
{d}s} >
0$ is
the expected number of true observations, $|E|$ denotes the Lebesgue
measure of $E$ and the mixing proportions $\pi_i$ sum up to one and
were set to $\pi_1 = 0.2$, $\pi_2 = 0.5$ and $\pi_3 = 0.3$. We consider
the sample sizes $\lambda_\mathrm{tot} = 1 000$, $\lambda_\mathrm{tot}
= 10\mbox{,} 000$ and $\lambda_\mathrm{tot} = 20\mbox{,} 000$,
which we refer to as the
small, medium and large sample size experiments, respectively. The true
space $E$ and the smeared space $F$ were both taken to be the interval
$[-7,7]$. The true points $X_i$ were smeared with additive Gaussian
noise of zero mean and unit variance. Points smeared beyond the
boundaries of $F$ were discarded from further analysis. With this
setup, the smeared intensity is given by the convolution
%
%
\begin{equation}
g(t) = (Kf) (t) = {\int_{E}^{} \mathcal{N}(t-s|0,1)f(s) \,\mathrm
{d}s},\qquad t \in F.
\label{eq:gmmSmearedIntensity}
\end{equation}
Note that this setup corresponds to the classically most difficult
class of deconvolution problems since the Gaussian error has a
supersmooth probability density [\citet{Meister2009}].

The smeared space $F$ was discretized using $n=40$ histogram bins of
uniform size, while the true space $E$ was discretized using order-4
B-splines with $L = 26$ uniformly placed interior knots, resulting in
$p = L + 4 = 30$ unknown basis coefficients. With these choices, the
condition number of the smearing matrix $\mathbf{K}$ was
$\operatorname{cond}(\mathbf
{K}) \approx2.6 \cdot10^8$, indicating that the problem is severely
ill-posed. The boundary hyperparameters were set to $\gamma_\mathrm{L}
= \gamma_\mathrm{R} = 5$. All experiments reported in this paper were
implemented in {\sc Matlab} and the computations were carried out on a
desktop computer with a quad-core 2.7~GHz Intel Core i5 processor. The
outer bootstrap loop was parallelized to the four cores of this setup.

With the exception of the number of bias-correction iterations, all the
algorithmic parameters were fixed to the same values for the three
different sample sizes. The MCEM algorithm was started using the
initial hyperparameter $\delta^{(0)} = 1 \cdot10^{-5}$ and was run for
30 iterations. The MCMC sampler was started from the nonnegative
least-squares spline fit to the smeared data, that is, $\bolds{\beta
}_\mathrm{init} = \min_{\bolds{\beta} \geq0} \|\tilde{\mathbf
{K}}\bolds{\beta} -
\mathbf{y}\|_2^2$, where the elements of $\tilde{\mathbf{K}}$ are
given by
equation \eqref{eq:Kij} with the smearing kernel $k(t,s) = \delta
_0(t-s)$, where $\delta_0$ denotes the Dirac delta function. The
sampler was then used to obtain 1000 post-burn-in observations from
the posterior. The number of bias-correction iterations was set to
$N_\mathrm{BC} = 15$ for the small sample size experiment, while
$N_\mathrm{BC} = 5$ iterations was used with the medium and the large
sample size. In each case, $R_\mathrm{BC} = 10$ bootstrap observations
were used to obtain the bias estimates and the bias-corrected
percentile intervals were obtained using $R_\mathrm{UQ} = 200$
bootstrap~observations.

\subsection{Results} \label{sec:simResults}

\subsubsection{Unfolded intensities}

%
%
%
%
%

For each sample size, 30 MCEM iterations sufficed for convergence to a
stable point estimate of the hyperparameter~$\delta$, with faster
convergence for larger sample sizes. In each case, there was little
Monte Carlo variation in the hyperparameter estimates. The estimated
hyperparameters were $\hat{\delta} = 2.0 \cdot10^{-4}$, $\hat
{\delta}
= 8.3 \cdot10^{-7}$ and $\hat{\delta} = 2.8 \cdot10^{-7}$ in the
small, medium and large sample size cases, respectively. A figure
illustrating the convergence of the MCEM iteration is given in the
online supplement [\citet{Kuusela2015}, Figure~3(a)]. In each case, the
running time to obtain $\hat{\delta}$, and thence the point estimate
$\hat{f}$, was approximately 8 minutes, while the time it took to form
the confidence intervals was approximately 14 hours for the medium and
large sample sizes and 39 hours for the small sample size, where three
times as many bias-correction iterations were performed. While these
are fairly long computations on a desktop setup, it should be noted
that it is easy to further parallelize the bootstrap computations and
the running time could be cut down substantially on a modern massively
parallel cloud computing platform.

\begin{figure}
\centering
\begin{tabular}{@{}c@{}}

\includegraphics{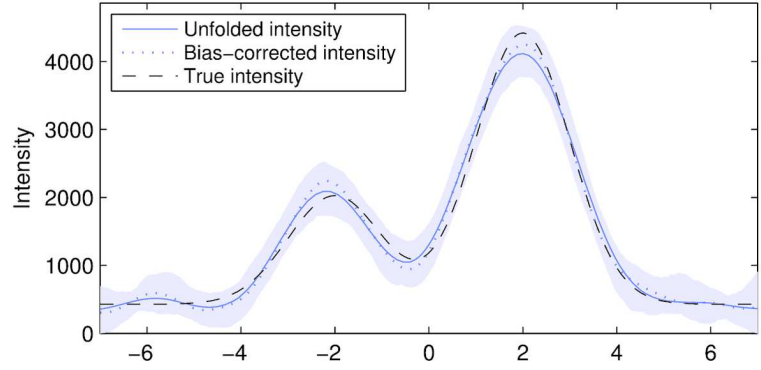}
\\
\footnotesize{(a) Gaussian mixture model data with $\lambda_\mathrm
{tot} = 20\mbox{,}000$}\\[3pt]

\includegraphics{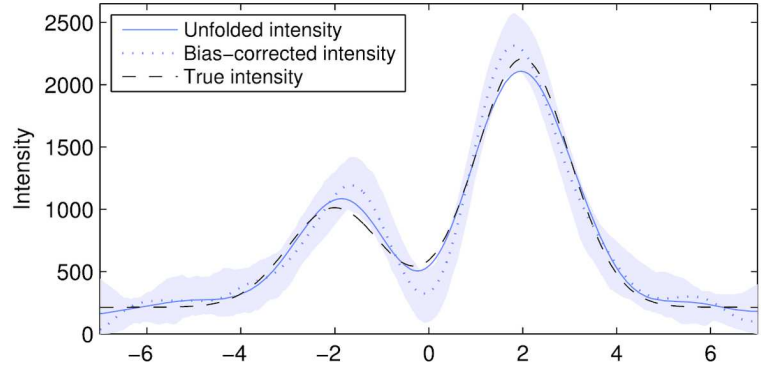}
\\
\footnotesize{(b) Gaussian mixture model data with $\lambda_\mathrm
{tot} = 10\mbox{,}000$}\\[3pt]

\includegraphics{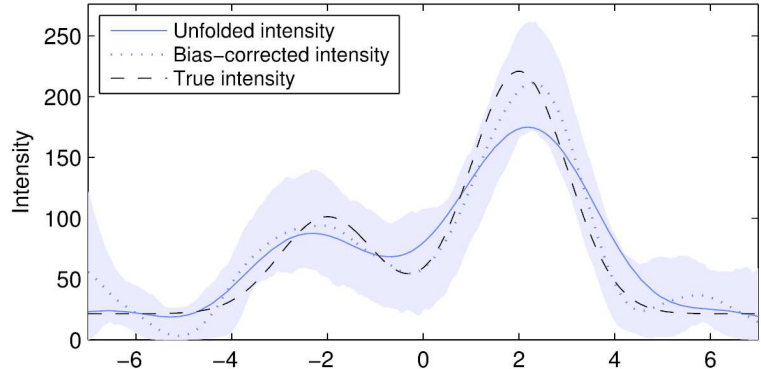}
\\
\footnotesize{(c) Gaussian mixture model data with $\lambda_\mathrm
{tot} = 1000$}
\end{tabular}
\caption{Unfolding results for the Gaussian mixture model data with \textup{(a)}
$\lambda_\mathrm{tot} = 20\mbox{,} 000$, \textup{(b)}~$\lambda_\mathrm{tot}
= 10\mbox{,} 000$
and \textup{(c)} $\lambda_\mathrm{tot} = 1000$. The confidence intervals are
the 95\% pointwise percentile intervals of the bias-corrected
estimator. The number of bias-correction iterations was 5 in cases \textup{(a)}
and \textup{(b)} and 15 in case \textup{(c)}.}
\label{fig:gmmUnfolded}
\end{figure}

Figure~\ref{fig:gmmUnfolded} shows the true intensities $f$, the
unfolded intensities $\hat{f}$ and the bias-corrected unfolded
intensities $\hat{f}_\mathrm{BC}$ for the three sample sizes. The
confidence bands consist of 95\% pointwise percentile intervals
obtained by bootstrapping the bias-corrected point estimate as
described in Section~\ref{sec:UQ}. In each case, the unfolded intensity
beautifully captures the two-humped shape of the true intensity, and,
unsurprisingly, the more data is available, the better the quality of
the estimate. It is also apparent that the estimators are biased near
the peaks and the trough of the true intensity, and the relative size
of this bias is larger for smaller sample sizes. In each case, the bias
can be reduced using the iterative bias-correction procedure, but at
the cost of an increased variance of the point estimator.

%

For each sample size, the confidence intervals perform well at covering
the true intensity without being excessively long or wiggly. For these
particular realizations, the intervals cover the truth at every point
$s \in E$ in the large and small sample size cases, while with
the medium sample size the intervals cover everywhere except for a
short section near $s = {-1.5}$ and another one near $s = 0$.

\subsubsection{Uncertainty quantification performance} \label{sec:UQperf}

We study the coverage performance of the bias-corrected intervals shown
in Figure~\ref{fig:gmmUnfolded} using a Gaussian approximation to the
Poisson likelihood described in Section~2.2 of the online supplement
[\citet{Kuusela2015}]. With this approximation, finding of the point
estimator $\hat{\bolds{\beta}}$ effectively reduces to a ridge regression
(i.e.,~Tikhonov regularization) problem for which the solution can be
computed in a fraction of the time required for the posterior mean with
the full Poisson likelihood. This enables us to perform an otherwise
computationally infeasible empirical coverage study for the
bias-corrected intervals. The Gaussian approximated intervals look
similar to the full intervals (see Section~3.2 of the supplement [\citet
{Kuusela2015}] for a detailed comparison), and we expect the coverage of
the full intervals to be similar to or better than that of the Gaussian
intervals.

The coverage of the bias-corrected intervals is compared to alternative
bootstrap and empirical Bayes intervals. The empirical Bayes intervals
we consider are the 95\% equal-tailed credible intervals induced by
the empirical Bayes posterior $p(\bolds{\beta}|\mathbf{y},\hat
{\delta})$. The
bootstrap intervals that we compare with are the standard percentile
intervals without bias-correction, which are obtained from our
procedure by setting the number of bias-correction iterations
$N_\mathrm
{BC}$ to zero. We also consider the bootstrap basic intervals which for
point $s \in E$ are given by $[2\hat{f}(s) - \hat
{f}_{0.975}^*(s),2\hat
{f}(s) - \hat{f}_{0.025}^*(s)]$, where $\hat{f}_{\alpha}^*(s)$ is the
$\alpha$-quantile of the bootstrap sample of nonbias-corrected unfolded
intensities obtained using a fixed hyperparameter~$\hat{\delta}$ and
the resampling scheme $\mathbf{y}^{*(1)}, \mathbf{y}^{*(2)}, \ldots,
\mathbf
{y}^{*(R_\mathrm{UQ})} \overset{\mathrm{i.i.d.}}{\sim}
\operatorname
{Poisson}(\mathbf{K}\hat{\bolds{\beta}})$ with $R_\mathrm{UQ} =
200$. The
bootstrap intervals are formed using the Gaussian approximation, while
the empirical Bayes intervals are for the full Poisson likelihood.

%
\begin{figure}
\centering
\begin{tabular}{@{}c@{}}

\includegraphics{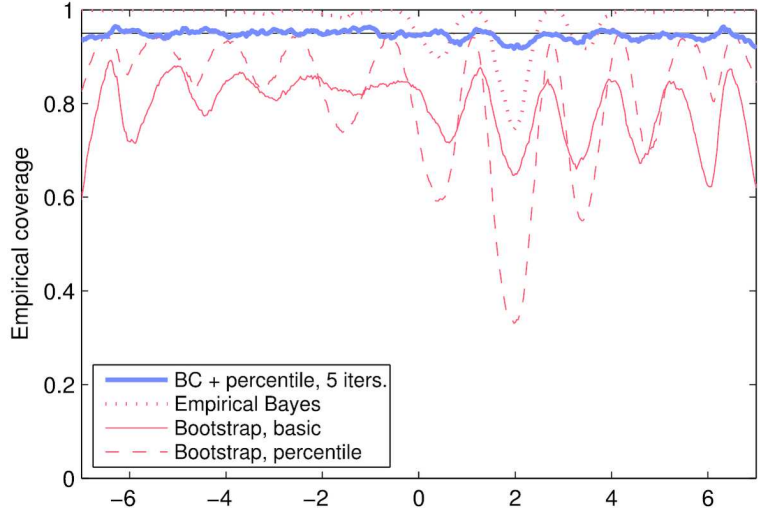}
\\
\footnotesize{(a) Comparison of UQ performance, $\lambda_\mathrm
{tot} =10\mbox{,}000$}\\[3pt]

\includegraphics{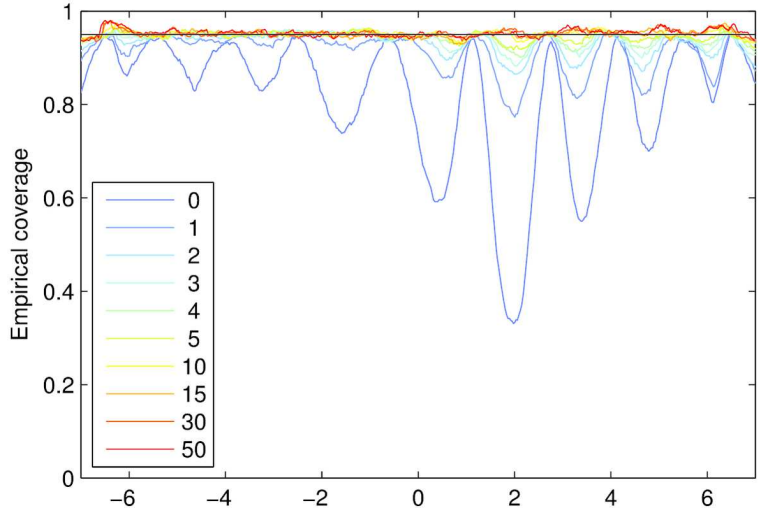}
\\
\footnotesize{(b) Effect of bias-correction on UQ performance,
$\lambda_\mathrm{tot} =10\mbox{,}000$}
\end{tabular}
\caption{Coverage studies for the Gaussian mixture model data with
expected sample size $\lambda_\mathrm{tot} =10\mbox{,} 000$.
Figure~\textup{(a)} compares the empirical coverage of the iteratively
bias-corrected intervals with 5 bias-correction iterations to that of
empirical Bayes credible intervals as well as the nonbias-corrected
bootstrap percentile and basic intervals. Figure~\textup{(b)} shows the
empirical coverage of the bias-corrected intervals as the number of
bias-correction iterations is varied between 0 and~50. All intervals
are formed for 95\% nominal coverage shown by the horizontal line.}
\label{fig:gmm10000Coverage}
\end{figure}

%

Figure~\ref{fig:gmm10000Coverage}(a) shows the empirical
coverage of the bias-corrected intervals obtained using the Gaussian
approximation and the various alternatives in the medium sample size
case for 1000 repeated observations from the smeared process $N$. The
coverage of the bias-corrected intervals is close to the nominal value
of 95\% throughout the spectrum (average coverage 94.6\%), although a
minor reduction in coverage due to the residual bias can be observed
near the peak at $s = 2$, where the lowest coverage is 91.7\%. A very
different behavior is observed with the empirical Bayes intervals:
these intervals overcover for most parts of the spectrum, but at
regions of significant bias they undercover. In particular, near $s =
2$, the worst-case coverage of these intervals is 74.5\%. The standard
bootstrap intervals, on the other hand, consistently undercover
(average coverage 81.6\% for the percentile and 79.2\% for the basic
intervals) and the nonbias-corrected percentile intervals in particular
fail to cover near the peak at $s = 2$, with empirical coverage of just
33.1\%.

To gain further insight into the improvement in coverage provided by
the bias-correction, we plot in Figure~\ref{fig:gmm10000Coverage}(b) the
empirical coverage for various numbers of bias-correction iterations.
For $N_\mathrm{BC} = 0$, the coverage is simply that of the standard
percentile intervals. We see that a single bias-correction iteration
already improves the coverage significantly, with further iterations
always improving the performance. In fact, increasing $N_\mathrm{BC}$
from 5, we are able to do away with the dip in Figure~\ref{fig:gmm10000Coverage}(a) around $s = 2$. However, we preferred
settling with $N_\mathrm{BC} = 5$, as increasing the number of
iterations produced increasingly wiggly intervals. Interestingly, the
coverage of the bias-corrected intervals does not seem to suffer from
the fact that all the bootstrap computations were performed without
resampling the hyperparameter estimate~$\hat{\delta}$, which suggests
that the uncertainty of $\hat{\delta}$ plays only a minor role in the
overall uncertainty of $\hat{f}$.

The coverage patterns reported here for $\lambda_\mathrm{tot} =
10\mbox{,} 000$
repeat themselves in the small and large sample size situations, with
starker differences between the methods with the small sample size and
milder differences with the large sample size. In the large sample size
case, the bias-corrected intervals effectively attain nominal coverage
(average coverage 94.7\%), with no major deviations from the nominal
due to residual bias. The bootstrap intervals, on the other hand,
undercover, while the empirical Bayes intervals overcover, except for
the peak at $s=2$ where slight undercoverage is observed. In the most
difficult small sample size case, both the bootstrap and the empirical
Bayes intervals have major problems with undercoverage, while the
bias-corrected intervals achieve close-to-nominal coverage on most
parts of the spectrum except for some undercoverage at $s = 2$, where
the worst-case coverage is 85.0\%. See the supplement [\citet
{Kuusela2015}] for detailed coverage plots for the small and large
sample size experiments as well as plots showing a realization of the
intervals for the different methods and numbers of bias-correction
iterations. The supplement also includes a comparison with hierarchical
Bayes intervals which are observed to provide qualitatively similar
coverage performance as the empirical Bayes intervals, although with
improved worst-case coverage for some choices of the hyperprior.

\section{Unfolding of the $Z$ boson invariant mass spectrum} \label
{sec:Zboson}

\subsection{Description of the data} \label{sec:ZbosonIntro}

In this section we illustrate the proposed unfolding framework using
real data from the CMS experiment at the Large Hadron Collider. In
particular, we unfold the $Z$ boson invariant mass spectrum published
in \citet{CMS2013ECAL}. The $Z$~boson, which is produced in copious
quantities at the LHC, is a mediator of the weak interaction. The
particle is very short-lived and decays almost instantly into other
elementary particles. The decay mode considered here is the decay of a
$Z$ boson into a positron and an electron, $Z \rightarrow e^+ e^-$. The
original purpose of these data was to calibrate and measure the
resolution of the CMS electromagnetic calorimeter, but they also serve
as an excellent testbed for unfolding since the true mass spectrum of
the $Z$ boson is known with great precision from previous measurements.

The electron and the positron produced in the decay of the $Z$ boson
are first detected in the CMS silicon tracker after which their
energies~$\mathcal{E}_i, i=1,2$, are measured by stopping the
particles at the ECAL; see Section~\ref{sec:physicsDetectors}. From
this information, one can compute the \emph{invariant mass} $W$ of the
electron--positron system defined by the equation
%
%
\begin{equation}
W^2 = (\mathcal{E}_1 + \mathcal{E}_2)^2
- \|\mathbf{p}_1 + \mathbf{p}_2\|_2^2,
\end{equation}
where $\mathbf{p}_i, i=1,2$, are the momenta of the two particles and the
equation is written using the natural units where the speed of light
$c=1$. Since $\|\mathbf{p}_i\|_2^2 = \mathcal{E}_i^2 - m_e^2$,
where $m_e$ is the rest mass of the electron, one can reconstruct the
invariant mass $W$ using only the ECAL energy deposits $\mathcal{E}_i$
and the opening angle between the two tracks in the silicon tracker.

The invariant mass $W$ is preserved in particle decays. Furthermore, it
is invariant under Lorentz transformations and has therefore the same
value in every frame of reference. This means that the invariant mass
of the $Z$~boson, which is simply its rest mass $m$, is equal to the
invariant mass of the electron--positron system, $W = m$. It follows
that measurement of the invariant mass spectrum of the
electron--positron pair enables us to measure the mass spectrum of the
$Z$ boson itself.

Due to the time-energy uncertainty principle, the $Z$ boson does not
have a unique rest mass $m$. Instead, the mass follows the Cauchy
distribution, also known in particle physics as the \emph{Breit--Wigner
distribution}, with density
%
%
\begin{equation}
p(m) = \frac{1}{2\pi} \frac{\Gamma}{(m-m_Z)^2 + {\Gamma
^2}/{4}}, \label{eq:BreitWigner}
\end{equation}
where $m_Z = 91.1876\ \mathrm{GeV}$ is the mode of the distribution
(often simply called \emph{the} mass of the $Z$ boson) and $\Gamma=
2.4952\ \mathrm{GeV}$ is the full width of the distribution at half
maximum [\citet{Beringer2012}]. Since the contribution of background
processes to the electron--positron channel near the $Z$ peak is
negligible [\citet{CMS2013ECAL}], the underlying true intensity $f(m)$
is proportional to $p(m)$.

The dominant source of smearing in measuring the $Z$ boson invariant
mass~$m$ is the measurement of the energy deposits $\mathcal{E}_i$ in
the ECAL. The resolution of these energy deposits is in principle
described by equation~\eqref{eq:ECAL_res}. However, when working on a
small enough invariant mass interval around the $Z$~peak, one can in
practice assume that the response is given by a convolution with a
fixed-width Gaussian kernel. Moreover, the left tail of the kernel is
typically replaced with a more slowly decaying tail function in order
to account for energy losses in the ECAL. It is therefore customary to
model the response using the so-called \emph{Crystal Ball \emph{(CB)}
function} [\citet{Oreglia1980,CMS2013ECAL}] given by
%
%
\begin{eqnarray}
&&\operatorname{CB}\bigl(m|\Delta m, \sigma^2, \alpha, \gamma\bigr)
\nonumber
\\[-8pt]
\\[-8pt]
\nonumber
&&\qquad =
\cases{ C e^{-{(m-\Delta m)^2}/{(2\sigma^2)}}, &\quad $\displaystyle\frac{m-\Delta
m}{\sigma} > -\alpha$,
\vspace*{2pt}
\cr
\displaystyle C \biggl( \frac{\gamma}{\alpha} \biggr)^\gamma
e^{-{\alpha^2}/{2}} \biggl( \frac{\gamma}{\alpha} - \alpha-
\frac{m - \Delta m}{\sigma}
\biggr)^{-\gamma}, &\quad $\displaystyle\frac{m - \Delta m}{\sigma} \leq- \alpha$,}
\end{eqnarray}
where $\sigma, \alpha> 0$, $\gamma> 1$ and $C$ is a normalization
constant chosen so that the function is a probability density. This
function is a Gaussian density with mean $\Delta m$ and variance
$\sigma
^2$, where the left tail is replaced with a power-law function. The
parameter $\alpha$ controls the location of the transition from
exponential decay into power-law decay and the parameter $\gamma$
controls the decay rate of the power-law tail. The forward mapping $K$
in equation \eqref{eq:FredIntEq} is then given by a convolution where
the kernel is the CB function, that is, $k(t,s) = k(t-s) =
\operatorname
{CB}(t-s|\Delta m, \sigma^2, \alpha, \gamma)$.

The data set we use is a digitized version of the lower left-hand plot
of Figure~11 in \citet{CMS2013ECAL}. These data correspond to an
integrated luminosity\footnote{The number of particle reactions that
took place in the accelerator is proportional to the integrated
luminosity. As such, it is a measure of the amount of data produced by
the accelerator. It is measured in units of inverse femtobarns,
$\mathrm
{fb}^{-1}$.} of $4.98\ \mathrm{fb}^{-1}$ collected at the LHC in 2011
at the $7\ \mathrm{TeV}$ center-of-mass energy and include 67 778
electron--positron events with the measured invariant mass between
65~GeV and 115~GeV. The data are discretized using a histogram with 100
bins of uniform width.
For details on the event selection, see \citet{CMS2013ECAL} and the
references therein.

In order to estimate the parameters of the Crystal Ball function, we
divided the data set into two independent samples by drawing a
binomial random variable independently for each bin with the number of
trials equal to the observed bin contents. Consequently, the bins of
the resulting two histograms are marginally mutually independent and
Poisson distributed. Each observed event had a 70\% probability of
belonging to the sample $\mathbf{y}$ used for the unfolding demonstration
and a 30\% probability of belonging to the sample used for CB
parameter estimation.

The CB parameters $(\Delta m, \sigma^2, \alpha, \gamma)$ were estimated
using maximum likelihood with the subsampled data on the full invariant
mass range 65--115~GeV assuming that the true intensity is proportional
to the Breit--Wigner distribution of equation~\eqref{eq:BreitWigner}.
The unknown proportionality constant of the true intensity was also
estimated as part of the maximum likelihood fit. The maximum likelihood
estimates of the CB parameters were
%
%
\begin{equation}
\bigl(\Delta\hat{m}, \hat{\sigma}^2, \hat{\alpha}, \hat{\gamma}
\bigr) = \bigl(0.56\ \mathrm{GeV}, (1.01\ \mathrm{GeV})^2, 1.95, 1.40
\bigr),
\end{equation}
indicating that the measured invariant mass is on average 0.56 GeV too
high and has an experimental resolution of approximately 1 GeV. As a
cross-check of the fit, the estimated Crystal Ball function was used to
smear the true intensity to obtain the corresponding expected smeared
histogram, which was found to be in good agreement with the observations.

\subsection{Unfolding setup and results}

To carry out the unfolding of the $Z$ boson invariant mass, we used the
subsampled $n = 30$ bins on the interval $F = [82.5\ \mathrm
{GeV},97.5\
\mathrm{GeV}]$. The resulting histogram $\mathbf{y}$ had a total of
42,475
electron--positron events. To account for events that are smeared
into the observed interval $F$ from the outside, we let the
true space $E = [81.5\ \mathrm{GeV}, 98.5\ \mathrm{GeV}]$,
that is, we extended it by approximately $1\hat{\sigma}$ on both sides
with respect to $F$. The true space $E$ was discretized using order-4
B-splines with $L = 34$ uniformly placed interior knots, resulting in
$p = 38$ unknown spline coefficients. It was found out that such
overparameterization with $p > n$ facilitated the mixing of the MCMC
sampler. With these choices, the condition number of the smearing
matrix was $\operatorname{cond}(\mathbf{K}) \approx8.1 \cdot10^3$.
The boundary
hyperparameters were set to $\gamma_\mathrm{L} = \gamma_\mathrm{R}
= 50$.

%
\begin{figure}

\includegraphics{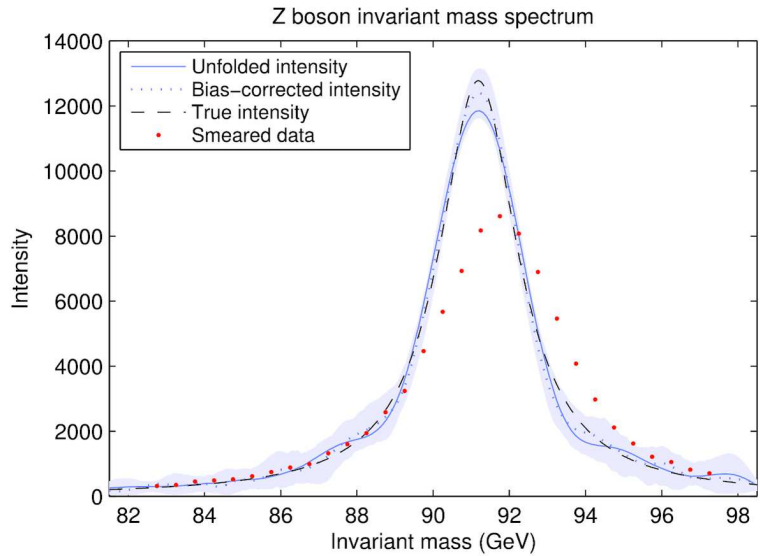}

\caption{Unfolding of the $Z$ boson invariant mass spectrum. The
confidence band consists of the 95\% pointwise percentile intervals of
the bias-corrected estimator obtained using 5~bias-correction
iterations. The points show a histogram estimate of the smeared
intensity.}
\label{fig:Zee}
\end{figure}

The parameters of the unfolding algorithm were set to the same values
as in the medium and large sample size experiments of Section~\ref
{sec:simulations}. In particular, the number of bias-correction
iterations was set to $N_\mathrm{BC} = 5$. The MCEM algorithm converged
after approximately 10 iterations to the hyperparameter estimate~$\hat
{\delta} = 7.0\cdot10^{-8}$ with little Monte Carlo variation. Finding
the hyperparameter $\hat{\delta}$ followed by the point estimate
$\hat
{f}$ took 10 minutes, while the running time to obtain the
bias-corrected confidence intervals was approximately 17 hours.

In Figure~\ref{fig:Zee}, the unfolded intensity $\hat{f}$ and the 95\%
pointwise percentile intervals of the bias-corrected intensity $\hat
{f}_\mathrm{BC}$ are compared with the Breit--Wigner shape of the $Z$
boson mass peak. The proportionality constant of the true intensity was
obtained from the maximum likelihood fit described in Section~\ref
{sec:ZbosonIntro}. The figure also shows a histogram estimate of the
smeared intensity given by the observed event counts $\mathbf{y}$ divided
by the 0.5 GeV bin width. The unfolding algorithm is able to correctly
reconstruct the location and width of the $Z$ mass peak which are both
distorted by the smearing in the ECAL. The intensity is also estimated
reasonably well in the 1 GeV regions in the tails of the intensity
where no smeared observations were available. More importantly, the 95\% bias-corrected confidence intervals cover the true mass peak across
the whole invariant mass spectrum. We also observe that the bias-correction is particularly important near the top of the mass peak,
where the original point estimate $\hat{f}$ appears to be somewhat~biased.

\section{Concluding remarks} \label{sec:discConc}

We have studied a novel approach to solving the high energy physics
unfolding problem involving empirical Bayes selection of the
regularization strength and frequentist uncertainty quantification
using an iterative bias-correction. We have shown that empirical Bayes
provides a straightforward, fully data-driven way of choosing the
hyperparameter~$\delta$ and demonstrated that the method provides good
estimates of the regularization strength with both simulated and real
data. Given the good performance of the approach, we anticipate
empirical Bayes methods to also be valuable in solving other inverse
problems beyond the unfolding~problem.

The natural fully Bayesian alternative to empirical Bayes consists of
using a Bayesian hierarchical model which necessitates the choice of a
hyperprior for $\delta$. Unfortunately, the resulting estimates are
known to be sensitive to this nontrivial choice [\citet{Gelman2006}]. We
provide in the online supplement [\citet{Kuusela2015}] a comparison of
empirical Bayes and hierarchical Bayes for a number of uninformative
hyperpriors. Our results indicate that the relative point estimation
performance of the methods depends strongly on the choice of the
hyperprior, especially when there is only a limited amount of data. For
example, in the small sample size experiment of Section~\ref
{sec:simulations}, the integrated squared error of hierarchical Bayes
ranges from 17\% better to 30\% worse than that of empirical Bayes
depending on the choice of the hyperprior. Unfortunately, there are no
guarantees that the uninformative hyperprior that worked the best in
this example would always provide the best performance. These results
also indicate that hyperpriors which were designed to be uninformative
are actually fairly informative in hierarchical Bayes unfolding and
have an undesirably large influence on the unfolded results, while
empirical Bayes achieves comparable point estimation performance
without the need to make any arbitrary distributional assumptions on
$\delta$. The hierarchical Bayes credible intervals also suffer from
similar coverage issues as the empirical Bayes credible intervals.

The other main component of our approach is frequentist uncertainty
quantification using an iterative bias-correction technique. We have
shown that the bias-correction is crucial for establishing
close-to-nominal frequentist coverage and that, at least within the
context of our simulation study, the approach outperforms existing
techniques based on bootstrap resampling and Bayesian credible
intervals. The good performance of this approach raises many
interesting questions for future work. For example, there seems to be a
trade-off between the coverage and the length of the bias-corrected
intervals and this trade-off is governed by the number of
bias-correction iterations $N_\mathrm{BC}$. That is, it appears that
one is able to obtain consistently better coverage by running more
bias-correction iterations at the expense of increased interval length.
By studying the theoretical properties of the iteration, one might be
able to provide guidance on how to choose $N_\mathrm{BC}$ in such a way
that it optimizes the length of the intervals while maintaining the
coverage within a preset threshold of the nominal value.

In this work, we have focused on the most basic inferential tool in
nonparametric statistics, that is, we have aimed at building pointwise
confidence bands for the unknown intensity $f$ with good frequentist
coverage properties. These bands can be directly used for making
pointwise inferential statements, such as testing if the data are
consistent with some theoretical prediction of $f$ at a single point $s
\in E$ chosen before seeing the data. They also form the basis for
making more complicated inferences. For example, one can envisage using
the bands for testing whether the data are consistent with a
theoretical prediction at several locations or over the whole spectrum
by making a multiple testing correction. The appropriate form of the
inferential statement depends, however, on the original purpose of the
unfolding operation and each purpose listed in Section~\ref
{sec:physicsUnfolding} involves different types of inferential goals,
including the comparison and combination of several unfolded
measurements. Extensions of the tools studied in this paper to such
more complicated situations constitutes an important topic for future work.

It should also be noted that throughout this paper the confidence
intervals are computed for a fixed forward mapping $K$, while in
reality there is usually a systematic uncertainty associated with $K$
stemming from the fact that $K$ is typically estimated using either
simulations or some auxiliary measurements. In cases where the detector
response can be modeled using a parametric model, such as the $Z$ boson
example of Section~\ref{sec:Zboson}, the unknown parameters can be
estimated using maximum likelihood on data generated by a known true
intensity. More generally, however, one may need to consider
nonparametric estimates of $K$. Once an estimate of $K$ has been
obtained, it is likely that its uncertainty can be incorporated into
the outer bootstrap loop of the bias-corrected confidence intervals.
A~detailed study of these ideas will be the subject of future work by
the authors.

Finally, it should be pointed out that it is possible to find
situations where the proposed unfolding methodology will not yield good
reconstructions. This happens when the smoothness penalty, that is,
penalizing for large values of $\|f''\|_2^2$, is not the appropriate
way of regularizing the problem. For instance, if the true intensity
$f$ contains sharp peaks or rapid oscillations, the solution would
potentially be biased to an extent where the iterative bias-correction
would be unable to adequately improve the situation. In particular, the
penalty would be unlikely to adapt well to situations where the
magnitude of the second derivative varies significantly across the
spectrum. This is the case, for example, when the intensity contains
features at very different spatial scales or when it corresponds to a
steeply falling spectrum that varies over many orders of magnitude.
Naturally, in these cases, a more suitable choice of the family of
regularizing priors $\{p(\bolds{\beta}|\delta)\}_{\delta> 0}$
should fix
the problem. For example, in the case of a steeply falling spectrum,
one could consider penalizing for the second derivative of $\log f$
instead of $f$ or using a spatially adaptive smoothness penalty [\citet
{Pintore2006}]. These considerations highlight the fact that all the
inferences considered here are contingent on the chosen family of
regularizing priors and should be interpreted with this in mind.

\section*{Acknowledgments}

We wish to warmly thank Bob Cousins, Anthony Davison, Tommaso Dorigo,
Louis Lyons and Mikko Voutilainen for insightful discussions and their
encouragement in the course of this work. We are also grateful to the
Associate Editor and the two anonymous reviewers for their
exceptionally detailed and constructive feedback.

\begin{supplement}[id=suppA]
\stitle{Supplement to ``Statistical unfolding of elementary particle
spectra: Empirical Bayes estimation and bias-corrected uncertainty
quantification''}
\slink[doi]{10.1214/15-AOAS857SUPP} 
\sdatatype{.pdf}
\sfilename{aoas857\_supp.pdf}
\sdescription{The supplement provides a comparison of the empirical
Bayes and hierarchical Bayes approaches to unfolding; additional
simulations results complementing those of Section~\ref
{sec:simulations}; and technical material on the convergence and mixing
of the MCMC sampler and on the Gaussian approximation used in the
coverage studies.}
\end{supplement}

%
%

%



\printaddresses
\end{document}